\documentclass[10pt,a4paper,twocolumn, twoside]{article}
\RequirePackage{graphicx}
\RequirePackage[switch]{lineno}
\RequirePackage{color}
\RequirePackage[colorlinks,citecolor=blue,urlcolor=blue,linkcolor=blue]{hyperref}
\usepackage{graphics}
\usepackage{graphicx} 
\usepackage{dcolumn}
\usepackage{bm}
\usepackage{multirow}
\usepackage{longtable}
\usepackage{booktabs}
\usepackage[latin1]{inputenc}
\usepackage{caption}
\usepackage[affil-it]{authblk}
\usepackage{blindtext}
\usepackage{abstract}
\usepackage{amsmath}

\makeatletter
\renewcommand{\section}{\@startsection{section}{2}{0cm}{-\baselineskip}
{0,5\baselineskip}{\normalsize\bfseries}}
\renewcommand{\subsection}{\@startsection{subsection}{3}{0cm}{-\baselineskip}
{0,5\baselineskip}{\normalsize\slshape}}
\makeatother

\begin{document}

\title{GIOVE -- A New Detector Setup for High Sensitivity Germanium Spectroscopy At Shallow Depth}

\author{G.~Heusser$\rm ^1$, M.~Weber$\rm ^1$\thanks{present address: Physics Department, Columbia University, New York, NY 10027, USA}, J.~Hakenm\"uller$\rm ^1$, M.~Laubenstein$\rm ^2$, M.~Lindner$ \rm ^1$, W.~Maneschg$\rm ^1$, H.~Simgen$\rm ^1$, D.~Stolzenburg$\rm ^1$, H.~Strecker$\rm ^1$}

\date{\small \it 
$^1$Max-Planck-Institut f\"ur Kernphysik, Saupfercheckweg 1, 69117 Heidelberg, Germany \\
$^2$Laboratori Nazionali del Gran Sasso, Via G. Acitelli 22, 67100 Assergi (AQ), Italy
\vspace{0.3cm}
\\ E-mail addresses: \\ {\tt Gerd.Heusser@mpi-hd.mpg.de \\
 Marc.Weber@astro.columbia.edu
} }
\vspace{0.3cm}

\twocolumn[
\begin{@twocolumnfalse}
\maketitle

\begin{abstract}
We report on the development and construction of the high-purity germanium spectrometer setup GIOVE (\underline{G}ermanium \underline{I}nner \underline{O}uter \underline{V}eto), recently built and now operated at the shallow underground laboratory of the Max-Planck-Institut f\"ur Kernphysik, Heidelberg. Particular attention was paid to the design of a novel passive and active shield, aiming at efficient rejection of environmental and muon induced radiation backgrounds. The achieved sensitivity level of $\leq$100 $\mu$Bq\,kg$^{-1}$ for primordial radionuclides from U and Th in typical $\gamma$ ray sample screening measurements is unique among instruments located at comparably shallow depths and can compete with instruments at far deeper underground sites. \\ \\
\noindent {\it Keywords:} low background gamma ray spectroscopy, low radioactive material screening, radiation shield, cosmic muon veto, neutron attenuation, radon suppression, Monte Carlo simulation \\
\end{abstract}
\end{@twocolumnfalse}
]
\vspace{1.0cm}

\saythanks

\section{Introduction}\label{sec:introduction}

\subsection{Low-level germanium spectroscopy}
Low-level $\gamma$ spectroscopy with germanium (Ge) detectors has become an essential tool for material screening in rare event physics experiments. These demand lowest possible radioactivity concentrations near their target or detector array~\cite{brodzinski1988,heusser1995}. Typical examples are searches for solar neutrinos, neutrinoless double beta decay and dark matter.\\Amongst others, there is the double beta decay experiment GERDA~\cite{ackermann2013} and the dark matter search project XENON~\cite{xenoninstrument}, both including participation of the Max-Planck-Institut f\"ur Kernphysik (MPIK) in Heidelberg, Germany.

Compared to other methods, such as mass spectrometry or neutron activation, Ge spectroscopy provides a more comprehensive method of material screening by collecting information about all contributing radioisotopes in a single energy spectrum. As a further advantage, material screening can be performed in a nondestructive way without complex sample treatment. Due to their environmental origin the primordial radioisotopes $^{232}$Th, $^{238}$U and $^{40}$K represent the most prominent sources of contamination in common materials. Concerning the two former isotopes, only Ge spectroscopy is capable of measuring the concentration of the most intensely $\gamma$ ray emitting progenies near the end of their respective decay chains -- in particular $^{208}$Tl and $^{214}$Bi. This enables a unique method to determine whether the secular equilibrium is broken or not. It also provides information about the current state of equilibrium by comparing the concentrations of the chain progenies $^{228}$Ac (representing its parent $^{232}$Th) and $^{234m}$Pa (representing the $^{238}$U chain) to those of the previously mentioned nuclides. This is of particular importance e.g. in the case of $^{232}$Th, $^{228}$Ra, $^{228}$Ac and the $^{228}$Th sub-chains because the time scale for noticeable changes in activity ratios is well within the duration of a typical low-background experiment.

The best available low-level Ge spectrometers are operated deep underground and reach specific count rate sensitivities up to $10$\,$\mu$Bq\,kg$^{-1}$~\cite{heusser2006}. This, however, requires long counting periods of up to 100 days. Consequently, several spectrometers must be run simultaneously in order to serve the needs of various experiments. The mentioned sensitivity is achieved by four detectors, named GeMPI, which were developed by MPIK and are operated at the LNGS underground facility in Assergi, Italy, below $\sim$3800~meters of water equivalent (m~w.e.)~\cite{heusser2006,rugel2005}. Besides them, there are three spectrometers of an earlier generation operated at the MPIK shallow underground laboratory ($15$\,m~w.e.)~\cite{aberle2010}. They are about $100$ times less sensitive than the GeMPI setups. The present GIOVE detector is designed to significantly lower the background at our near surface location by combining efficient measures of active and passive radiation shielding. By this means, we aim at bridging the gap between shallow depth and deep underground counting facilities.
	
\subsection{Design principles of GIOVE}\label{sec:designprinciples}

\begin{figure}[tb]
\includegraphics[width=1.0\columnwidth]{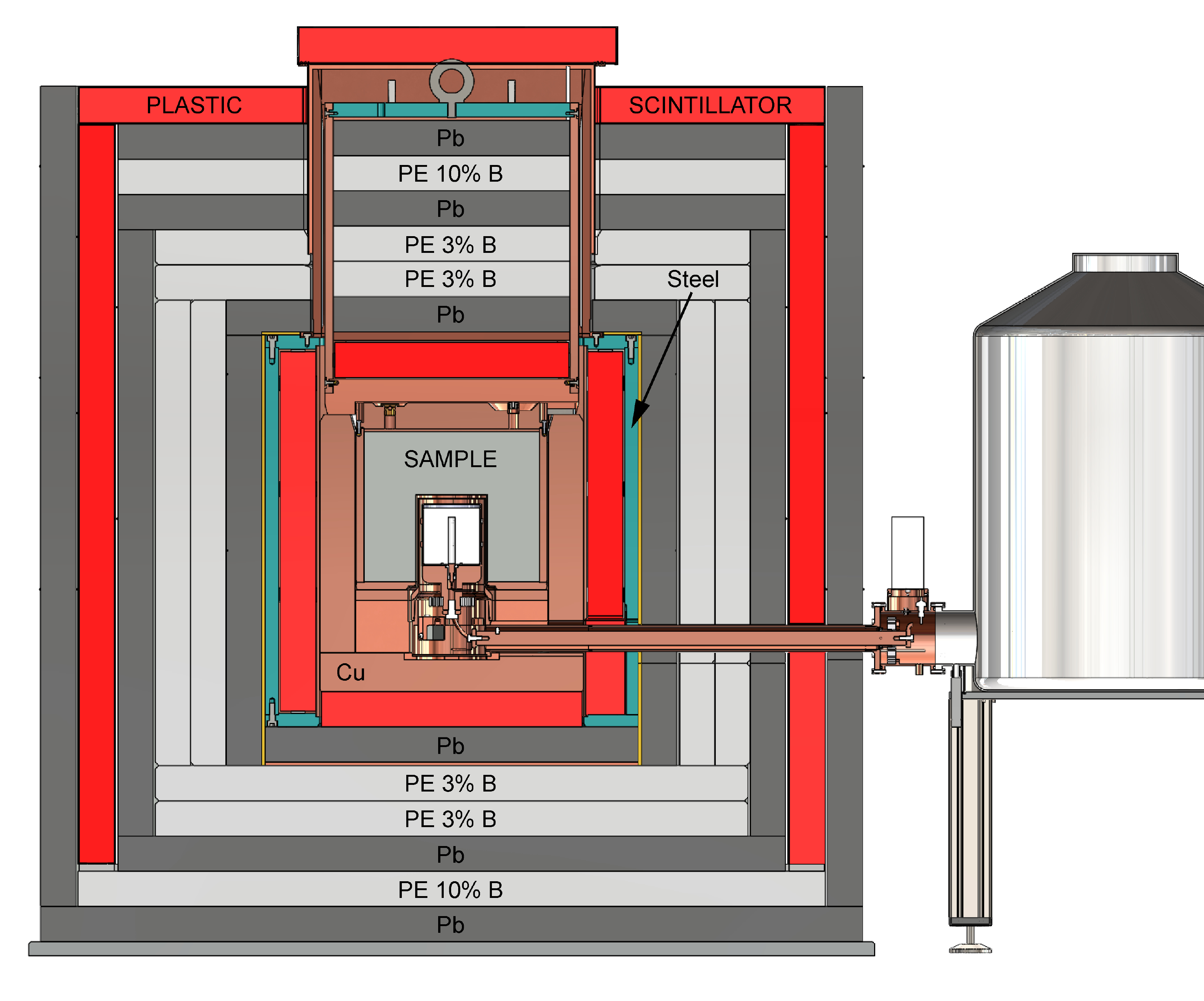}
\caption{Cut-away view of the GIOVE detector and shield. Different layers for radiation absorption (Pb, Cu), neutron moderation (PE) and active vetoing (red colored areas) form a shell structure to efficiently lower the background count rate in the central Ge crystal. More details provided in the text.}
\label{fig:gioveshield}
\end{figure}

The main objective of GIOVE is to enable unprecedented sensitivity in the $100\;\mu$Bq\,kg$^{-1}$ range in low-level radioactivity measurements of large screening objects at a shallow depth of only about $15$\,m~w.e. At this depth the muon flux is reduced only by a factor of $\sim$2$-$3~\cite{heusser1993}. Therefore, the aimed background suppression requires a cosmic muon veto efficiency of $\geq99\%$ as well as the ability to reduce the neutron induced background component as much as possible. The latter component consists mainly of delayed de-excitation lines of the isomeric states $^{71m}$Ge, $^{73m}$Ge and $^{75m}$Ge, which are produced by neutron capture on Ge inside the detector. Figure~\ref{fig:gioveshield} shows a schematic view of the multi-layer concept, designed to approach the reduction of the two mentioned background sources.

Plastic scintillator detectors are chosen for the muon veto system because they feature high photon yield and fast signal response. Moreover, they provide a similar neutron thermalization power as polyethylene (PE), which is often used for moderation (see also Section~\ref{sec:neutronattenuation}). We opted for a two layer structure of plastic scintillator plates. One layer aligned to the outer shield surface and the other surrounding the detector center. In this way muons escaping detection through outer edges and corners, or only causing an energy deposition below the sensitive threshold, have still a chance to be recorded by the inner plates. 

The outer veto system is supposed to suppress background events also from secondary processes which are induced by muons crossing the setup only peripherally. In total seven individual plastic scintillator plates are arranged such that they provide maximum angular coverage from all directions except the floor, which has to withstand the ton-scale weight of the entire shield. The lateral scintillator plates are placed behind the outermost layer of Pb in order to achieve an improved separation of recorded muon signals from the high energy tail of the environmental $\gamma$ ray spectrum and, thus, avoid sizable live time loss from an otherwise increased veto count rate. A further advantage of this alignment is that we expect the most efficient moderation of muon induced neutrons by placing the moderating materials directly behind the Pb layer, as predicted by initial Monte Carlo (MC) neutron simulations to find the optimal layer sequence. Due to mechanical constraints the same implementation was not applicable for the top part of the shield setup and plastic scintillator plates had to be left uncovered by additional Pb. 

Complementary, the inner veto consists of six individual plates, each $5$\,cm in thickness, together forming a cubic shell structure that encloses the innermost copper (Cu) chamber from all sides ($4\pi$~coverage). Placed inside the shield, underneath the Pb brick layers, an almost complete suppression of external $\gamma$ rays allows for a fairly low veto trigger threshold to achieve very high muon tagging efficiency. Due to the reduced trigger rate it is further possible to prolong the event-based veto hold-off to hundreds of $\mu$s and, hence, reject delayed de-excitations of neutron-induced Ge isomers even more efficiently.

Like any component close to the Ge detector the inner veto system has to be constructed of especially radio-pure materials, i.e. with low levels of intrinsic radioactivity, to avoid an increase of the total background level. Line background rates caused by primordial radionuclides below $1$~d$^{-1}$ in the Ge detector are pursued. This requires similar precaution as taken in the construction of the GeMPI detectors~\cite{heusser2006}. Special measures also relate to the manufacture of the cryostat system near the Ge crystal and inner shield, made of high purity Cu with only short cosmogenic sea level exposure. We also applied electron-beam welding wherever possible as well as electropolishing before the final assembly under clean room conditions. All other materials have been selected carefully after performing Ge~$\gamma$~ray screening.

The constraint to keep the target mass for muonic neutron production low requires that the total shield should be as compact as possible. However, it must remain sufficiently large to attenuate the intensity of the $2.6$\,MeV $\gamma$ line emitted from the laboratory walls. This resulted in three layers of $5$\,cm Pb, $4$ layers of $5$\,cm moderation material -- either of B loaded PE or of plastic scintillator -- and about $7$\,cm Cu plus $2$\,cm steel as supporting structure.

The radon (Rn) suppression system of GIOVE is configured such to avoid time delays at the beginning of measurements due to the decay of unequilibrated Rn or Rn progenies. Depending on the structure of the sample (solid, powder or liquid) and the decaying Rn isotope, the necessary enclosure time of the sample may last several hours ($^{222}$Rn-progenies $^{214}$Pb, $^{214}$Bi), days ($^{220}$Rn-progeny $^{212}$Pb) or even weeks ($^{222}$Rn) before any conclusive measurement can be taken. Therefore, each sample is placed into a gas-tight Cu container of ($249\times249\times216$)~mm$^3$ dimensions with a central cylindrical recess of \O$103$\,mm$\times124$\,mm to fit around the detector end cap (effective volume about $12.4$\,l). The containers are flushed with boil-off nitrogen (N$_2$), evaporating from an outside storage tank, before being sealed. Next, the container is transferred through a N$_2$ pressurized air lock into a glovebox system under slight overpressure. Here, it can be stored until the accumulated radionuclides from Rn decay on the container surface have completely decayed, or -- in the case of samples with high emanation rate -- until the equilibrium between $^{226}$Ra and $^{222}$Rn is re-established.

\section{Shield Design and Construction}\label{sec:shield}

\subsection{Inner and outer muon veto}\label{sec:muonveto}
An important specification for the implementation of an efficient muon veto system is a high signal yield which provides satisfying discrimination between $\gamma$ and muon signals and, thus, allows us to minimize the amount of dead-time caused by $\gamma$ rays and accidental triggers. This turns out to be particularly important in case of the external veto layers, which are more strongly exposed to environmental $\gamma$ ray activity. A sufficient mechanical and long-term stability is also required to ensure steady operation without need to open the passive shield once the construction is finished. These combined specifications resulted in our choice of the plastic scintillator type EJ-200.

Because of its approved low radiation background performance in the \textsc{XENON100} dark matter experiment~\cite{xenoninstrument} and its compact dimensions we employ \textsc{Hamamatsu} photomultiplier tubes (PMTs) R8520 as scintillation light detection devices in the inner veto system. These instruments feature an $1$\,in squared sensitive photocathode adapted to the emission wavelength of the applied plastic scintillator.

Before purchase and implementation of the inner veto system extensive studies have been performed in order to optimize the scintillation light collection efficiency (LCE) and its spatial homogeneity in the employed plastic scintillator plates. As a side constraint the number of PMTs, necessary for scintillation light read-out, is supposed to be small in order to maximize the sensitive target area of the veto. We approach the optimization problem by two complementary means:
\begin{enumerate}
\item A cosmic ray muon telescope setup was built to measure the scintillation light signal response depending on the spatial position where muons traverse the scintillator plates~\cite{stolzenburg2011}. A threefold coincidence condition is applied to constrain the area of muon incidence on the surface within a few cm$^2$: Simultaneous signal appearance is required in the actual scintillator volume and within two additional \O$2$\,in$\times2$\,in cylindric plastic scintillator detectors placed vertically above and below along a common axis. The muon induced light emission in the central detector is recorded with the mounted R8520 PMT and the coincidence spectrum is evaluated for various well-localized positions of the telescope.
\item A custom-made light propagation simulation, named \textsc{LuxIter}, has been developed to predict the expected number of photons arriving on a defined active surface after being isotropically emitted from a fixed spatial location or along a certain initial track~\cite{weber2009}. The program is capable of handling any custom-made geometry based on planar boundaries. The simulated photon tracing considers optical light absorption, parameterized by the specific light attenuation length at the maximum emission wavelength ($\lambda\approx380\,$cm for type EJ-200), and light reflection at boundaries, determined by the refractive indices of interfacing materials ($n=1.58$ for common plastic scintillators).
\end{enumerate}
The results of the LCE simulation are verified using data acquired with the muon telescope. Since the absolute photon yield of the measured muon signals is unknown we compare only relative signal amplitudes between different light emission positions. The predicted relative yields agree very well with measured data -- showing position dependent relative deviations of $<5\%$ for simple plate geometries and $<10\%$ for more complicated structures (non-rectangular edges and volume cuttings)~\cite{stolzenburg2011}. Very similar experimental agreement was found in further cross-checks using collimated $\gamma$ calibration sources placed on the scintillator surface. The consistency of these comparisons encouraged us to apply the LCE simulation for a study of various mounting configurations, e.g. different numbers of PMTs, their fixture and inclination angles. If only one single read-out PMT is considered the optimal configuration (highest photon gain and LCE uniformity) is obtained for a PMT placement inside a corner recess with its photo-cathode window pointing towards the center of the plate. The application of more than one PMT per scintillator plate is finally disfavored as the total LCE is already sufficient for setting an effective muon threshold and any additional device would increase mounting complexity, introduce further radio-impurities and reduce the effective area of the scintillator.

For the outer veto system we rely on commercially available, large area scintillator plates with integrated single PMT read-out, manufactured by Scionix Holland BV. The mounted PMT type ELT 9900 provides an enhanced 2$\pi$ photo-cathode sensitivity. The performance of the outer plates is likewise tested and approved for muon peak separation and signal uniformity before installation.

\begin{figure}[tb]
\includegraphics[width=1.0\columnwidth]{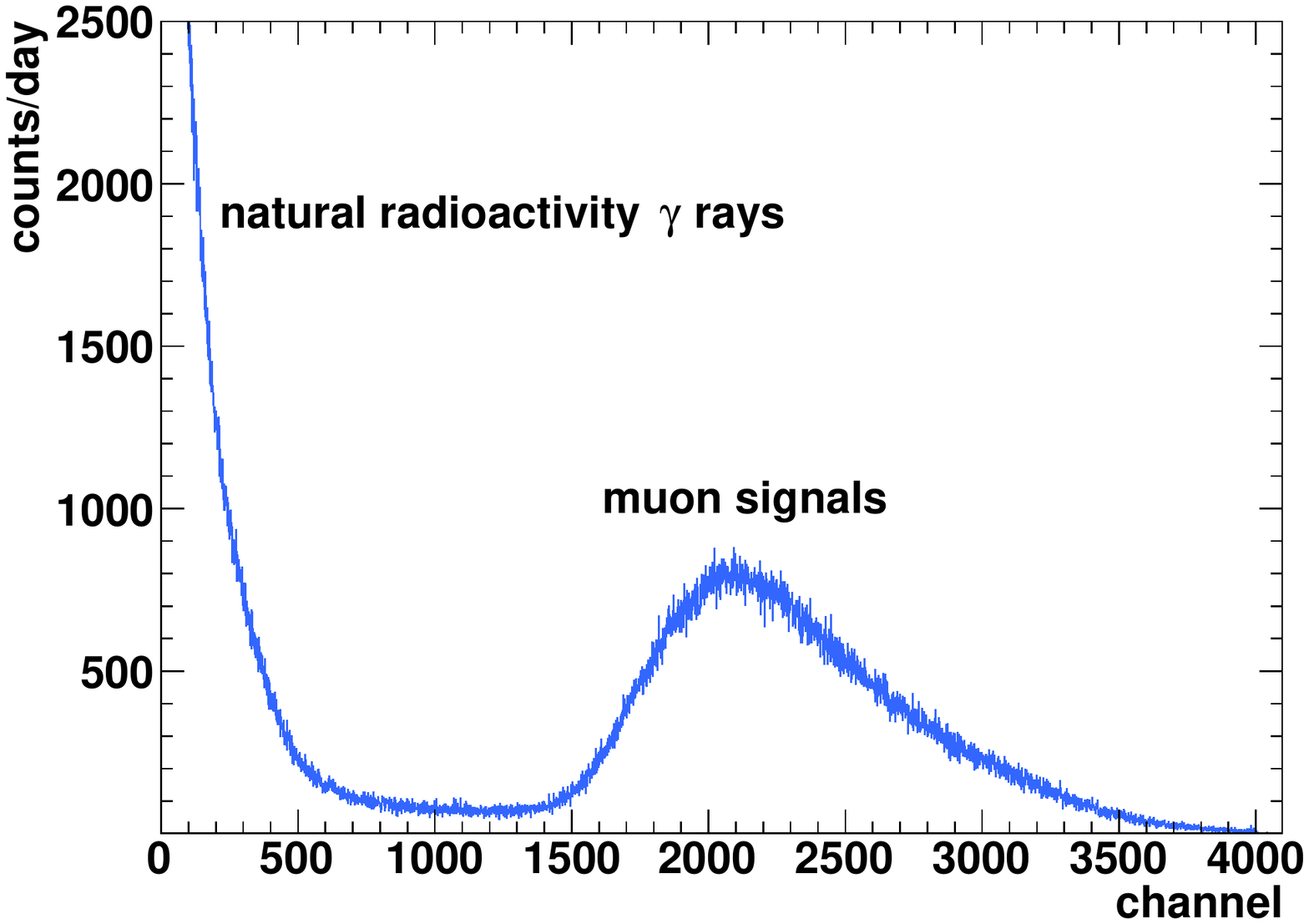}
\caption{A typical spectrum measured using one of the inner veto plastic scintillators underneath a $5$~cm Pb layer. The muon peak is clearly separated from the high energy tail of $\gamma$ rays from environmental radioactivity.}
\label{fig:scintillatorspectrum}
\end{figure}

The entire muon veto system is supposed to reject muon induced signals in the Ge detector with an efficiency $\geq99$\%. At the same time, the total veto trigger rate is required to remain low to minimize the accumulated Ge detector dead-time. The excellent separation power between the high energy tail of natural radioactivity $\gamma$ rays and the mean muon energy deposit is depicted by a typical pulse-height spectrum in Figure~\ref{fig:scintillatorspectrum}, taken by one of the plastic scintillator plates underneath 5\,cm of Pb shield. For each subsystem, the individual discriminator thresholds are defined such to constrain the rate of high energy $\gamma$ events, leaking into the muon acceptance range, to $\leq$2000\,d$^{-1}$. At that rate, the dead-time contribution from accidental $\gamma$ events is negligibly low.

In order to estimate the corresponding muon tagging efficiency as a function of the individual threshold setting, a dedicated coincidence setup with two scintillator plates on top of each other, separated by a $5$\,cm thick layer of Pb bricks, is installed. Simultaneous events in both scintillator layers are considered muon induced since the probability for random accidental $\gamma$ events becomes very small. By comparing the count rates of the tested scintillator with and without applying the chosen discriminator threshold, the muon acceptance can be calculated. As a result, all scintillators of the inner veto show muon acceptance fractions higher than $96\%$. The number provides a lower bound on the acceptance expectation for the completed shield setup, because some muon events in the test setup deposit only a small amount of energy by crossing only the edges of the scintillator. Such muon events, however, are likely to be identified by another scintillator within the $4\pi$ shield setup of GIOVE. 

For the outer veto a similar but simplified procedure is applied: For each scintillator plate, the trigger rate is measured as a function of an increasing discriminator threshold. At some value, the observed rate dropped notably, which is interpreted as crossing the endpoint of natural $\gamma$ radioactivity at 2.6~MeV. This set point was chosen to serve as the lower discrimination threshold for subsequent muon veto operation. Still exposed to a much higher trigger rate from the outer veto scintillators, an anti-coincidence gate window of $82$\,$\mu$s is applied to keep the summed dead-time contribution of the whole veto system below an aimed threshold of $\sim$2\%.

For the implementation of in total $13$ veto signal channels in the final GIOVE setup, programmable discriminator NIM modules were designed and custom-made at MPIK, and are used to generate one single veto gate signal. The boards are also capable of performing any logical combination of their input/output signals and can generate a final gate signal of adjustable length in time. The eventual performance of the muon veto system inside the completed detector shield will be part of our discussion in Section~\ref{subsection:giove-bckg-analysis}.

\subsection{Neutron attenuation} \label{sec:neutronattenuation}
Ambient neutrons contribute to background counts in the GIOVE detector via nucleon capture or (in)elastic scattering on the Ge crystal. While the primary neutron component of the cosmic radiation is considerably suppressed by the $\sim$15\,m~w.e. overburden above the shallow depth laboratory, there are two further production schemes: Neutrons from nuclear spallation, induced by cosmic muon interaction inside the passive detector shield and surrounding walls, and radiogenic neutrons from ($\alpha$,n) reactions or spontaneous fission of natural uranium (U) and thorium (Th) isotopes~\cite{heusser1993}. In order to stop these neutrons efficiently, compact neutron moderators and absorbers are needed. Compounds made of hydrogenous organic materials, such as PE and paraffin, and elements with a high thermal neutron capture cross section, such as boron (B) or lithium (Li), are widely used in nuclear physics experiments and radiation protection shields for this purpose.
\begin{figure}[tb]
\includegraphics[width=1.0\columnwidth]{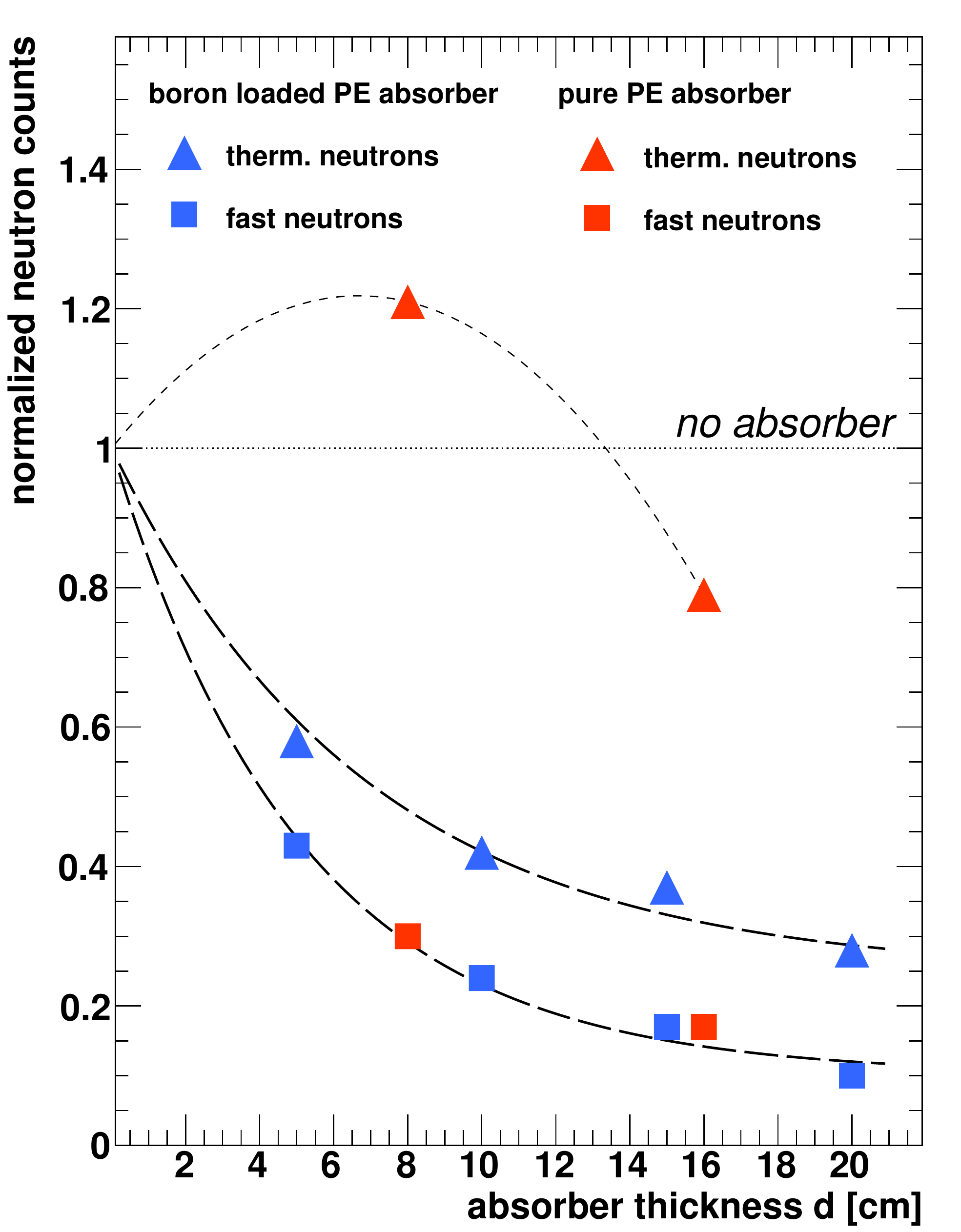}
\caption{Neutron attenuation with varying absorber thickness, relative to the no absorber measurement, as measured in the linear test setup (refer to text for details). We present results for pure and B loaded PE used as attenuation material. Long-dashed lines represent an exponential fit to the observed rate decline. The short-dashed line is for illustration only and represents a cubic-spline interpolation of the thermal neutron data for the pure PE absorber.}
\label{fig:neutronattenuation}
\end{figure}
To obtain a reliable quantification of the neutron flux attenuation through different pure and B loaded PE absorbers we perform a series of measurements in a dedicated neutron source tabletop experiment. Along a common line of sight we place a broadband $^{241}$AmBe neutron source ($\sim$$7.4\times10^{5}$\,s$^{-1}$), up to several layers of absorption material and a Ge counter to detect inelastic neutron events on top of the ambient $\gamma$ spectrum. The available absorptive plates had dimensions of about ($50\times30$)~cm$^2$ area and $5$\,cm/$8$\,cm thickness each, and the total distance between source and detector was about $45$\,cm. Additional Pb bricks around the source location are placed to shield from intense $\gamma$ ray activity of the $^{241}$Am decay. For each absorber configuration the signal strength of inelastic neutron reactions on the Ge target is measured. We separately analyze count rates from characteristic $\gamma$ rays emitted after the neutron capture reactions $^{72}$Ge($n,\gamma$)$^{73m}$Ge (53~keV) and $^{74}$Ge($n,\gamma$)$^{75m}$Ge (140~keV). We evaluate also the inelastic $^{72}$Ge($n,n'$)$^{73m}$Ge scattering process at $\sim$$691$\,keV. While the two former reactions feature large cross sections for thermal  neutrons at low kinetic energies, the latter process is only viable for relatively fast neutrons above the energy transfer threshold of the metastable $^{73m}$Ge. We make use of this property to distinguish between interactions of fast and slow neutrons on the Ge target.

Figure~\ref{fig:neutronattenuation} shows a summary of the most relevant results of the attenuation measurement. All count rates evaluated from the various inelastic neutron reactions are normalized to the case of no absorber placed. As expected, the relative flux of the fast neutron component decreases as a function of the PE absorber thickness because the initial kinetic energy spectrum is effectively shifted from MeV energies (the broad-band $^{241}$AmBe source spectrum has a mean energy of $\sim$5~MeV~\cite{klugeundweise1982}) into the thermal range. This explains why the relative rate of thermalized neutrons is first observed to rise above the zero absorption level if only pure PE plates are placed between source and detector. Otherwise, using B loaded ($5\%$) PE, the converted thermal neutrons are attenuated almost as efficiently as the high energetic ones.

In our setup, a complete suppression of the neutron flux is impossible because attenuation acts only within the solid angle covered by the absorber plates and cannot prevent neutrons from backscattering off ambient walls. We account for this constant background in our proposed fit model of the normalized number of neutron counts $n(d)$ as a function of absorber thickness $d$,
\begin{equation*}
n(d) = 1 - f\cdot(1 - \mathrm{e}^{-d/l})~.
\end{equation*}
Thereby, it is assumed that only the fraction $f$ of events passes through the actual absorption layers. We derive the mean attenuation length $l=(6.9\pm0.7)$\,cm and $l=(5.1\pm0.2)$\,cm for the thermal and fast neutron components, respectively, from the fitted data of the B loaded ($5\%$) PE, presented in Figure~\ref{fig:neutronattenuation} (best fit functions shown by long-dashed lines). In doing so, the fraction $f$ is allowed to float in the fit, amounting to $75\%$ ($89\%$) for the thermal (fast) measured neutron components. The small difference in $f$ is not surprising given the expected variations in background scattering patterns from the two classes of kinetic neutron energy.

We conclude that only the absorber loaded PE will help to diminish both the thermal and high energetic part of the neutron flux in the passive shield setup of GIOVE. Having determined that the mean absorption length is considerably less than $10$\,cm we decide that already three $5$\,cm layers of B loaded PE in every direction would provide a sufficient mitigation of the expected neutron induced background, thereby preventing the shield dimensions from becoming too bulky.

The here presented straight direction measurements are also confirmed by subsequent studies, adding layers of B loaded PE on top of an existing 4$\pi$ closed Pb shield construction, equipped with another Ge spectrometer inside. We deduce from this measurement that the sequential ordering of Pb and PE layers has no significant impact on the neutron attenuation power. This motivates our choice of having at least one layer of Pb shield be placed between the inner detection chamber of GIOVE and the innermost layer of PE to suppress any effect from radio-impurities in the neutron absorber (see Section \ref{sec:radiopurity}).
	
\subsection{Radio-purity and material selection for the Ge detector and shield} \label{sec:radiopurity}

The central part of the detector is a \O$80$\,mm$\times80$\,mm Ge diode of 2.1\,kg mass, which was produced in $2004$ with less than one week of sea level exposure in the framework of the GERDA~\cite{ackermann2013} experiment. It had been stored since then in the HADES underground facility in Mol, Belgium, until it was used for GIOVE.

We know from experience that hardly any material normally used in Ge detectors and shield construction is reliably radio-pure in the main primordial radionuclides at levels of $0.01$ to several mBq\,kg$^{-1}$. Only electrolytic Cu seems to be an exception. A~former Cu cryostat of the Heidelberg-Moscow experiment~\cite{guenther1997} has been refurbished for GIOVE. A new holder and end cap are machined from high purity Cu to fit around the crystal. The conical seal between the cryostat hat and the end cap is equipped with an O-ring to improve the long term vacuum stability. Like for the GeMPI detectors the cooled field-effect transistor (FET) assembly is shielded against the detector by a small brick made from low activity Pb (see also Figure~\ref{fig:simulation of cryostat}, right).

Our low background goal eventually requires all inner materials used near the Ge detector to be screened and to fulfill a high radio-purity level. The level should not exceed the $2.6$\,MeV external $\gamma$ ray line contribution after attenuation through the GIOVE Pb and Cu shield. Depending on the mass of the individual object the applied screening sensitivity ranges from some $10$\,$\mu$Bq\,kg$^{-1}$ (LNGS) to about $1$\,mBq\,kg$^{-1}$ (MPIK). The tolerable activity level is relaxed with increasing distance from the crystal. All samples undergo careful cleaning procedures after being machined, as e.g. described in Ref.~\cite{maneschg2008} for stainless steel. Plastic pieces are cleaned using only isopropanol. A selection of the most important screening results is presented in Table~\ref{tab:activitymeasurements} and is discussed below in sequence.

\begin{table*}[tb] \small
\caption[]{Radio-impurity concentration of various materials tested for the GIOVE setup. The upper four measurements were performed at LNGS using the GeMPI-$1$ detector. The lower three samples were screened using our existing low-level Ge detectors Bruno and Corrado at MPIK. Refer to Figure~\ref{fig:gioveshield} for indication at which position of the shield some of the investigated parts are located.}
		\label{tab:activitymeasurements}
		\centering
		\begin{tabular*}{\textwidth}[tbh]{@{\extracolsep{\fill}}lcccccc}
		\hline
		sample & amount & unit & \multicolumn{4}{c}{activity concentration [mBq\,unit$^{-1}$]} \\
		 & & & $^{226}$Ra & $^{228}$Th & $^{228}$Ra & $^{40}$K \\
		\hline
		plastic scintillator NE-102A & $7.30$ & kg & 0.96$\pm$0.13 & 9.3$\pm$0.3 & 9.2$\pm$0.4 & 2.2$\pm$0.7 \\
		plastic scintillator EJ-200 & $7.26$ & kg & 0.09$\pm$0.05  & 0.5$\pm$0.1 & 0.4$\pm$0.1 & 1.5$\pm$0.5 \\
		GALLEX carbon steel & $51.2$ & kg & 0.13$\pm$0.03 & $<$0.04  & 0.10$\pm$0.04 & 0.5$\pm$0.3 \\
		high-density PE (HDPE) & $7.76$ & kg & 0.27$\pm$0.08 & $<$0.14 & $<$0.50 & $<$3.2 \\
		PMT R8520-106 MOD                      & $7$    & piece &  0.36$\pm$0.21  &  $<$0.4  &  $<$0.5 &  7.4$\pm$1.7   \\
		B$_2$O$_3$ p.a. for silic. anal. Merck & $4.0$  & kg    &  $<$1.2         &  $<$1.9  &  $<$2.3 &  84$\pm$8      \\
		black Lexan scint. coverage            & $1.43$ & kg    &  11$\pm$2       &  $<$4    &  $<$10   &  116$\pm$18   \\
		\hline
		\end{tabular*}
\end{table*}

Two different types of plastic scintillators are measured as aliquot samples for the inner veto system and EJ-200 was found superior to NE-102A from an earlier NaI(Tl) spectrometer~\cite{heusser1986}. A steel frame of the former NaI(Tl) system was foreseen as supporting structure for the heavy shielding block. Fortunately, its screening resulted in one of the most radio-pure steel samples we had ever measured~\cite{maneschg2008}. Originally, this steel batch was selected for the GALLEX shielding tank~\cite{heusser1994}, which explains the labeling. Like in the former shield setup the steel frame holds the inner plastic scintillator plates. Small items of the cryostat, like the insulation beaker directly around the crystal or the sliding guidance pads used for the shielding block and the sample chamber, are made from high-density PE (HDPE, commercial name TECAFINE PE~$10$). Also the B loaded moderation plates are manufactured from HDPE. Only $^{226}$Ra can be identified as a positive signal in this HDPE sample. Slightly positive results in some of the tested radioisotope concentrations of the PMT R$8520$-$106$ Mod are tolerable due to the fact that for the inner veto system only one of them is used per scintillator plate (Section~\ref{sec:muonveto}). 

A major effort remarked the search for B or Li compounds which satisfy the radio-purity expectations and which are chemically stable against the elevated temperature of the PE polymerization process. Li or B admixed to the PE are considered because of their high cross-section for thermal neutron capture with a low $\gamma$ emission yield at the same time (see also Section~\ref{sec:neutronattenuation}). Many measurements of various, mainly B loaded compounds, and also tests of PE polymerization with admixture of these at the company \textsc{Profilan} have finally resulted in the boron oxide (B$_2$O$_3$) listed in Table~\ref{tab:activitymeasurements}. It still shows a rather high $^{40}$K concentration. This prevented us from the original plan of placing $5$\,cm borated PE around the inner copper chamber to shield against neutrons produced also in the innermost $5$\,cm Pb layer. As depicted in Figure~\ref{fig:gioveshield}, two layers of borated ($3\%$) PE have to be separated from the Ge detector by $5$\,cm of Pb to compensate for the high $^{40}$K concentration level of the admixed B$_2$O$_3$. The optimal B concentration (also to save costs) is evaluated with moderation measurements performed with the Ge spectrometer Corrado~\cite{budjas2008}, which is also operated in the MPIK underground facility. The relative intensity of $^{71m}$Ge, $^{73m}$Ge and $^{75m}$Ge, excited by residual neutrons, is measured for PE samples with variable B concentrations of $1$, $3$, $5$ and $10$\%. The samples fully occupy the sample chamber of about $12.4$\,l. Compared to our eventual choice of $3$\% admixture there is little or no further reduction using $5$ or $10$\% B, respectively. However, with respect to $1$\%, the signal rate still decreased by about a factor of $1.5-2$. In the same way, the moderation effect of pure PE and plastic scintillator material is compared and found to be similar within statistics.

Black Lexan with a thickness of $0.45$\,mm is used for the lightproof wrapping of the scintillator plates, together with black tape. Taking into account its overall low mass contribution, the $^{40}$K activity is comparable to that of the scintillator itself.
	
\subsection{Glove-box system and shield construction} \label{sec:shieldconstruction}

\begin{figure}[tb]
\includegraphics[width=1.0\columnwidth]{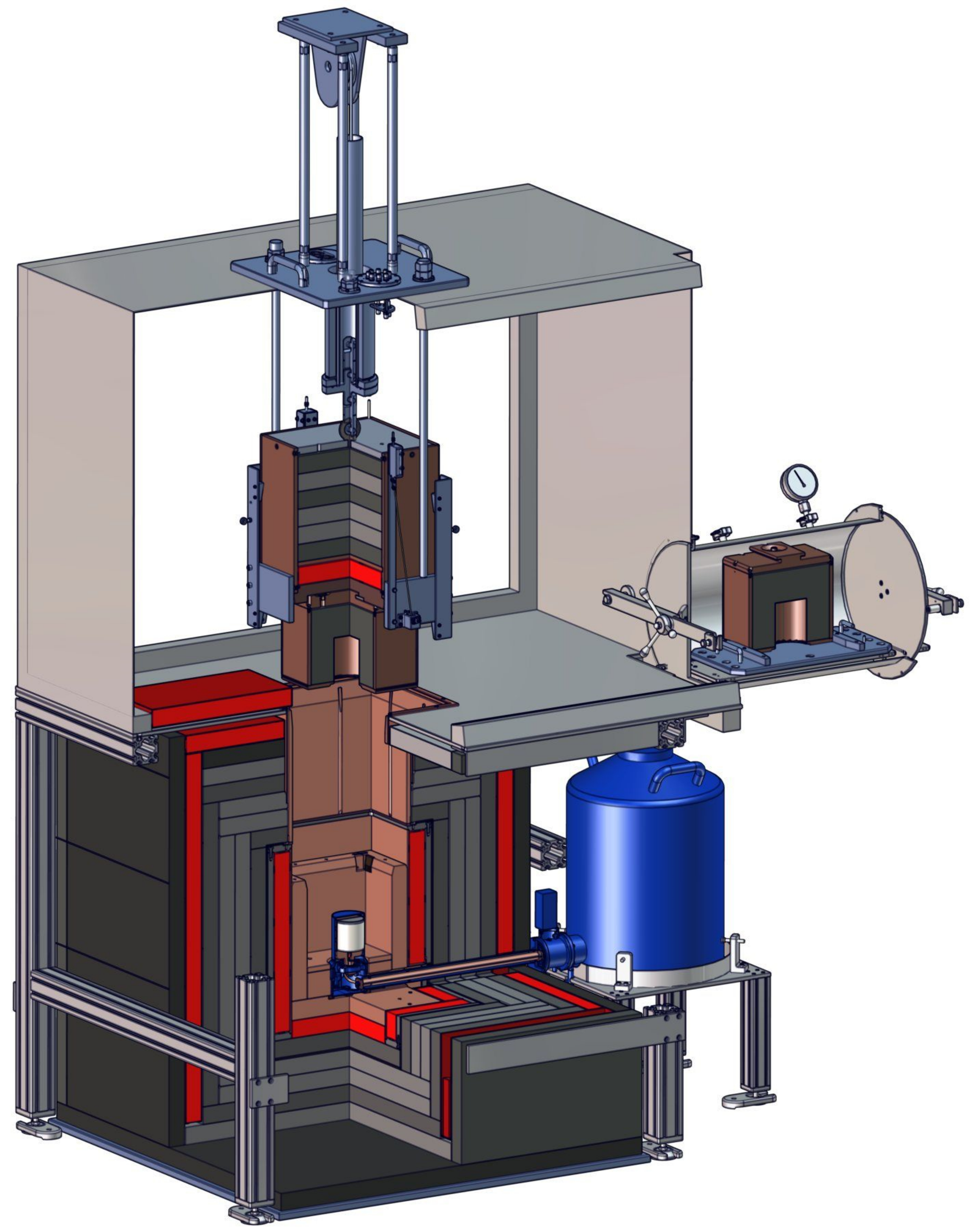}
\caption{Schematic of the entire GIOVE setup with one sample chamber mounted underneath the shield block and another one inside the antechamber. The total height is $2.45$\,m, including the glove-box.}
\label{fig:giovesetup}
\end{figure}

\begin{figure*}[t]
\begin{minipage}[t]{0.62\linewidth}
\begin{center}
\includegraphics[width=0.6\linewidth]{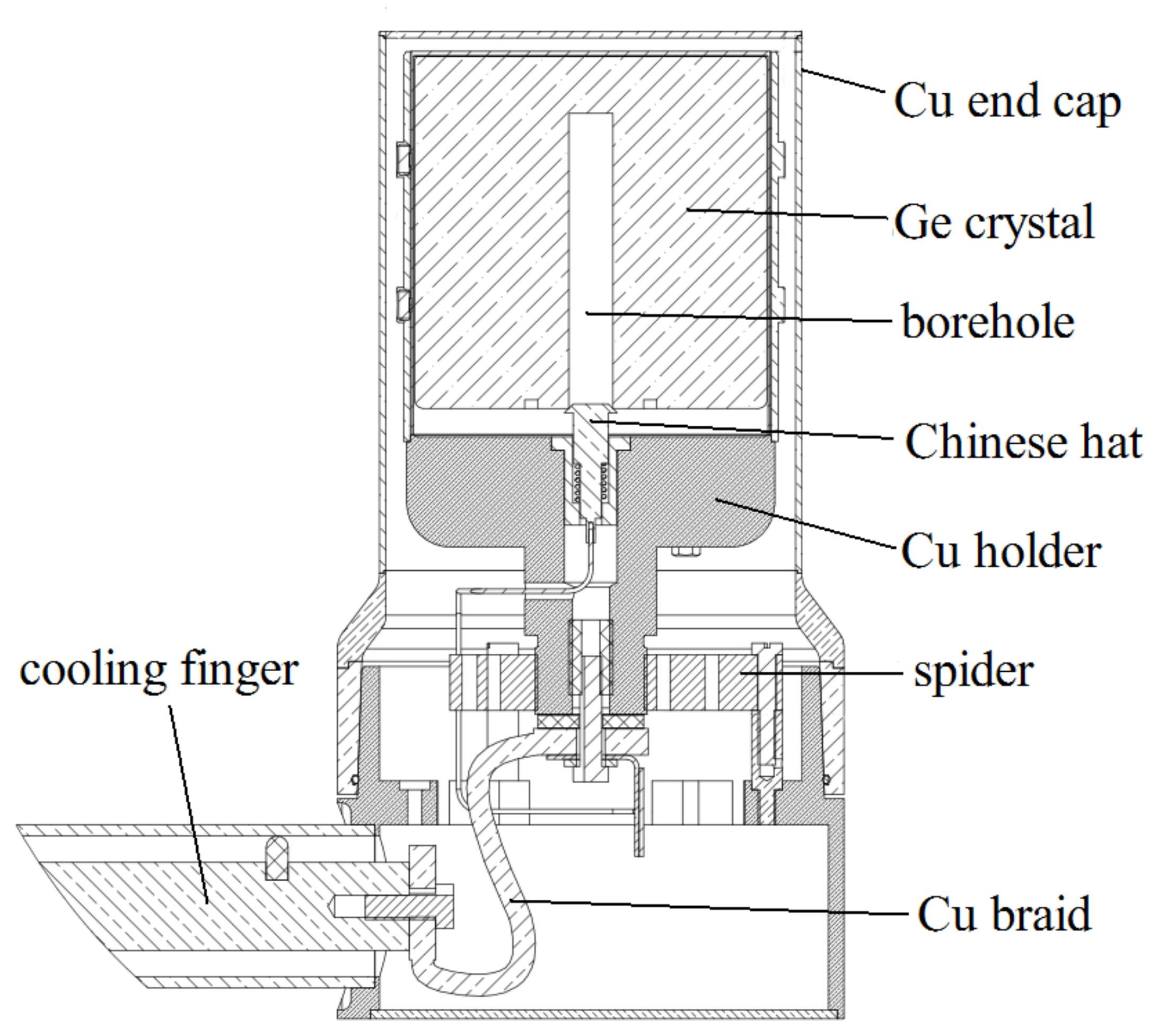}
\end{center}
\end{minipage}
\hfill
\begin{minipage}[t]{0.41\linewidth}
\includegraphics[width=0.6\linewidth]{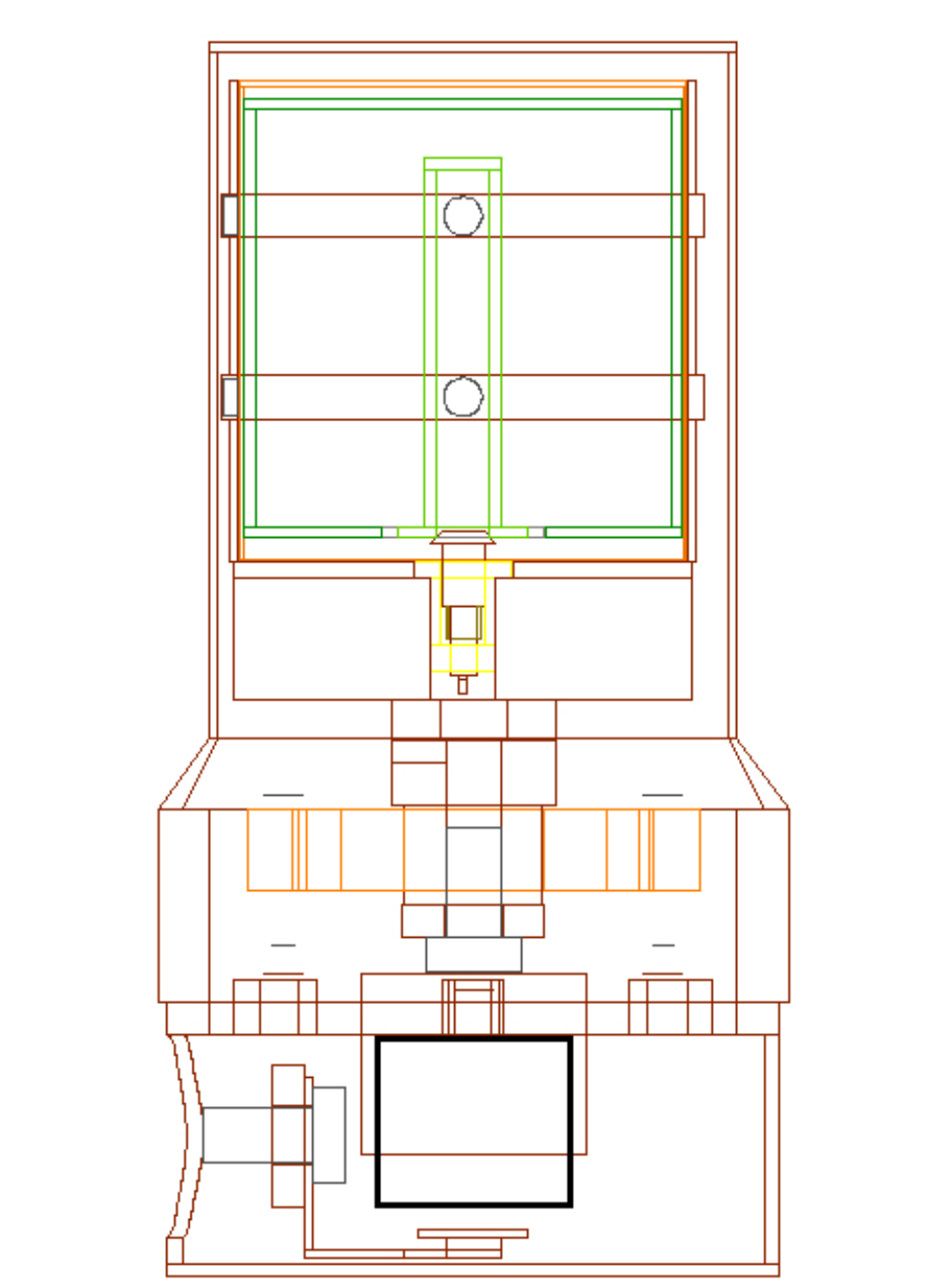}
\label{l}
\end{minipage}
\caption{High detail model of the Ge diode in the vacuum cryostat end cap. Left: construction plan, right: graphic representation of the simulated setup geometry. Here the small Pb block (black line) to shield the Ge crystal against the FET assembly is depicted in addition.}
\label{fig:simulation of cryostat}
\end{figure*}

The sequence of moderator plates and plastic scintillators enlarges the shield in such a way that the sample is not directly accessible from outside, as it is the case for the GeMPI detectors. This has affected the design of the shield and of the glove-box system on top of it. Figure~\ref{fig:giovesetup} shows the entire setup. 
A sample is typically placed outside of the system in a Cu container, resting on a motion platform with ballcasters mounted underneath. Next, the container is transferred through the antechamber into the glove-box. After being turned by $90^{\circ}$ its conical top fits into the counter part of the lifted shield block. In this locked position the motion platform can be removed and the sample chamber (still connected to the shielding block) can be lowered into the final counting position. This motion is performed through a rope and controlled by a ceiling mounted spindle-drive. Finally, the topmost veto counter is slid over the lowered shield block -- as shown in Figure~\ref{fig:gioveshield} -- and the measurement can be started.

Inside the glove-box there is storage space for all four constructed sample containers, which are sitting on their motion platforms. In the radon free N$_2$ atmosphere long term protection against plate-out activity and dust is assured. Moreover, Rn, which is incorporated in the sample or could not be removed during the original N$_2$ purging of the container, as well as plated out Rn progenies, continue to decay (see also Section~\ref{sec:designprinciples}).

The N$_2$ purge gas enters through a Teflon tube, which is installed below the cryostat arm and ends directly under the detector head. Thus, the gas flow permanently vents through the sample chamber into the glove-box volume. An additional N$_2$ supply on top of the glove-box can be turned on, if necessary. Surplus N$_2$ leaves the box through a bubbler vessel and ensures to maintain an overpressure of about $2-5$\,mbar. Using the same sealed canal underneath the cryostat a second tube enables the insertion of a wire mounted calibration source (typically $^{228}$Th). Also this tube ends below the detector head at a fixed position, enabling occasional calibrations to test the detector energy, resolution and efficiency stability.

All parts used in the shield construction, e.g. every Pb brick, have been at least wipe-cleaned with isopropanol. Every device within the inner veto frame is additionally treated in ultrasonic bath, cleaned with acid or, if possible, electropolished. The latter is generally applied for all inner Cu parts, including the cryostat components next to the crystal. The electropolishing also included the $11$\,mm thick Cu plates and the central $0.3$\,mm Cu caps of the four sample containers before their assembly using electron-beam or laser welding. Each of the containers has a Marinelli-like geometry and is able to carry up to $150$\,kg on top of its own weight of about $35$\,kg.
 
The Cu used for GIOVE was purchased together with other experiments as one batch of type ``NOSV'' (higher than $99.99$\% Cu, except gaseous impurities, e.g. up to $0.04$\% of O$_2$) from the company \textsc{Aurubis}, Germany, and was hot formed into the required dimensions and partly machined immediately after the electrolysis at the company \textsc{Carl Schreiber}, Germany. To guarantee a low cosmogenic activation level~\cite{cebrian2010,heusser2009} the Cu used has always been stored underground in the MPIK underground laboratory whenever it was not needed above ground for machining and cleaning.

\section{Detector Characterization and Modelling}\label{sec:detectormodelling}
\subsection{Geant4 detector model}
\label{sec:mcmodel}
The detector and shield geometry has been implemented in a full Monte Carlo (MC) simulation using the Geant$4$-based toolkit MaGe~\cite{boswell2011}. MaGe has been developed jointly by the GERDA and MAJORANA collaborations in order to adapt simulations to low-energy particle physics. The detector geometry has been implemented according to the construction plans and measured dimensions of the setup components. The vacuum cryostat containing the diode (Figure~\ref{fig:simulation of cryostat}) and the entire shield up to the outermost Pb layer have been implemented in detail~\cite{hakenmueller2012}. Surface scans allowed us to identify the actual position and active volume dimensions of the Ge crystal with high accuracy even after it had been mounted underneath the cryostat end cap (see Section~\ref{sec:Surface scan and dead-layer measurements}). The detector simulation is supposed to serve several purposes:

First, it is used in the routine detector application, most prominently at determining the collection efficiency of $\gamma$ rays from screened samples. The efficiency depends dominantly on placement, geometry and self-absorption attributes of the sample as well as of the surrounding materials. All of this is taken into account by the simulation. In order to keep the simulation errors on the efficiency as small as possible, several parameters (dead-layer thickness, active volume, diode position) in the simulation need to be adjusted to match real properties. The corresponding parameters are evaluated in a comparative study, combining calibration measurements with simulated data.

Second, the simulation is useful to locate the position of $\gamma$ emitting impurities by comparing data and MC generated spectra of radioactive sources placed at different parts of the setup.

Third, in future simulations the propagation and impact of muons and neutrons which penetrate the shield and detector will be studied to enhance our current understanding of these processes and their effect on the measured background in the GIOVE detector.
 	
\subsection{Surface scan and dead-layer measurements}
\label{sec:Surface scan and dead-layer measurements}
In order to achieve reliable and accurate predictions of $\gamma$ ray detection efficiencies using MC simulations it is crucial to model the details of the Ge crystal, including its precise dead-layer thickness and its exact placement inside the cryostat end cap.

\begin{figure*}[t]
\begin{minipage}[t]{0.5\textwidth}
\includegraphics[width=1.0\textwidth]{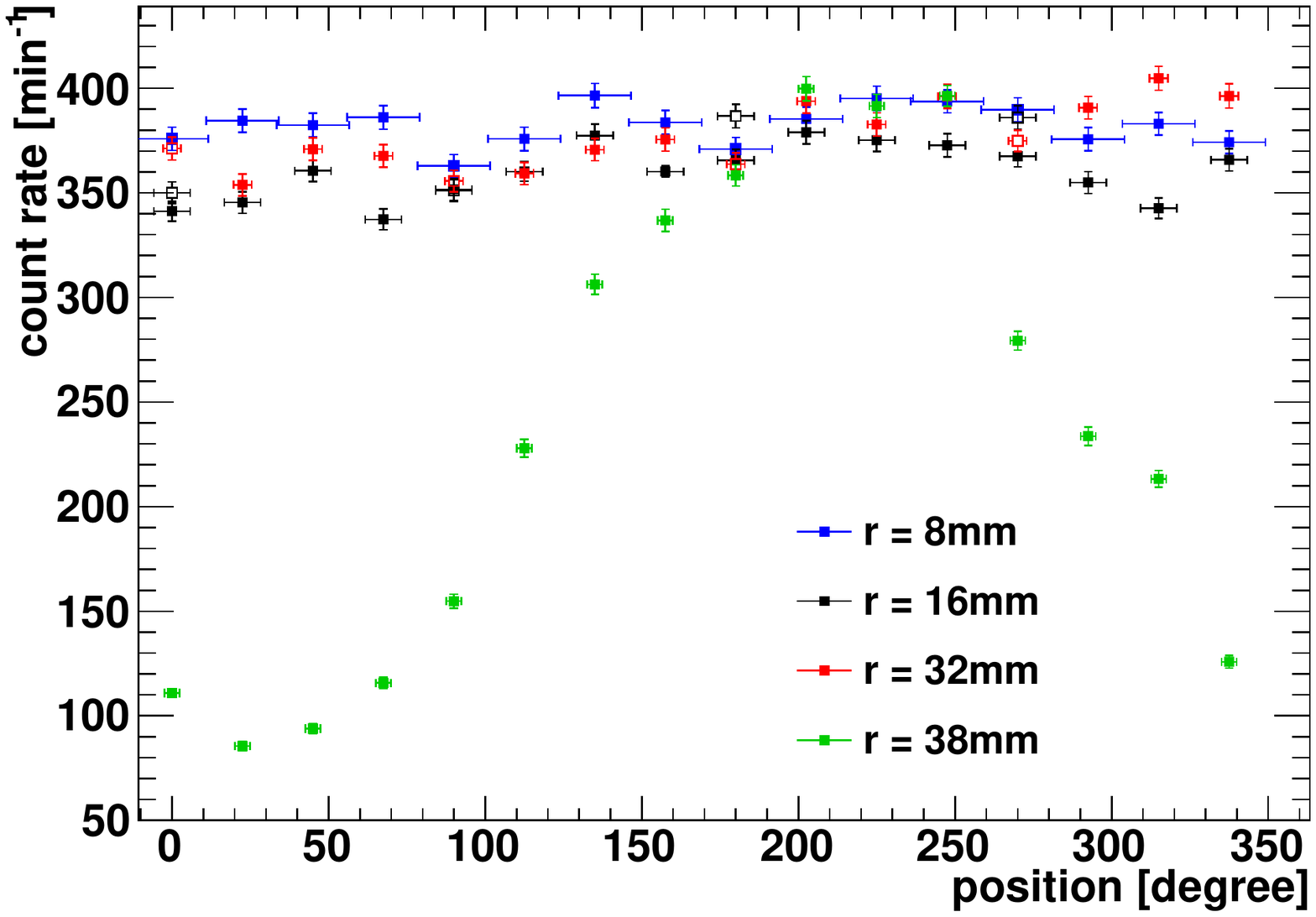}
\caption{Result from $^{241}\rm Am$ surface scans of the GIOVE crystal to determine the dead-layer homogeneity along circles on top of the diode. Different colors correspond to scans performed at different radii~\cite{hakenmueller2012}.}
\label{fig:circle scans}
\end{minipage}
\hfill
\begin{minipage}[t]{0.48\textwidth}
\begin{center}
\includegraphics[width=0.82\linewidth]{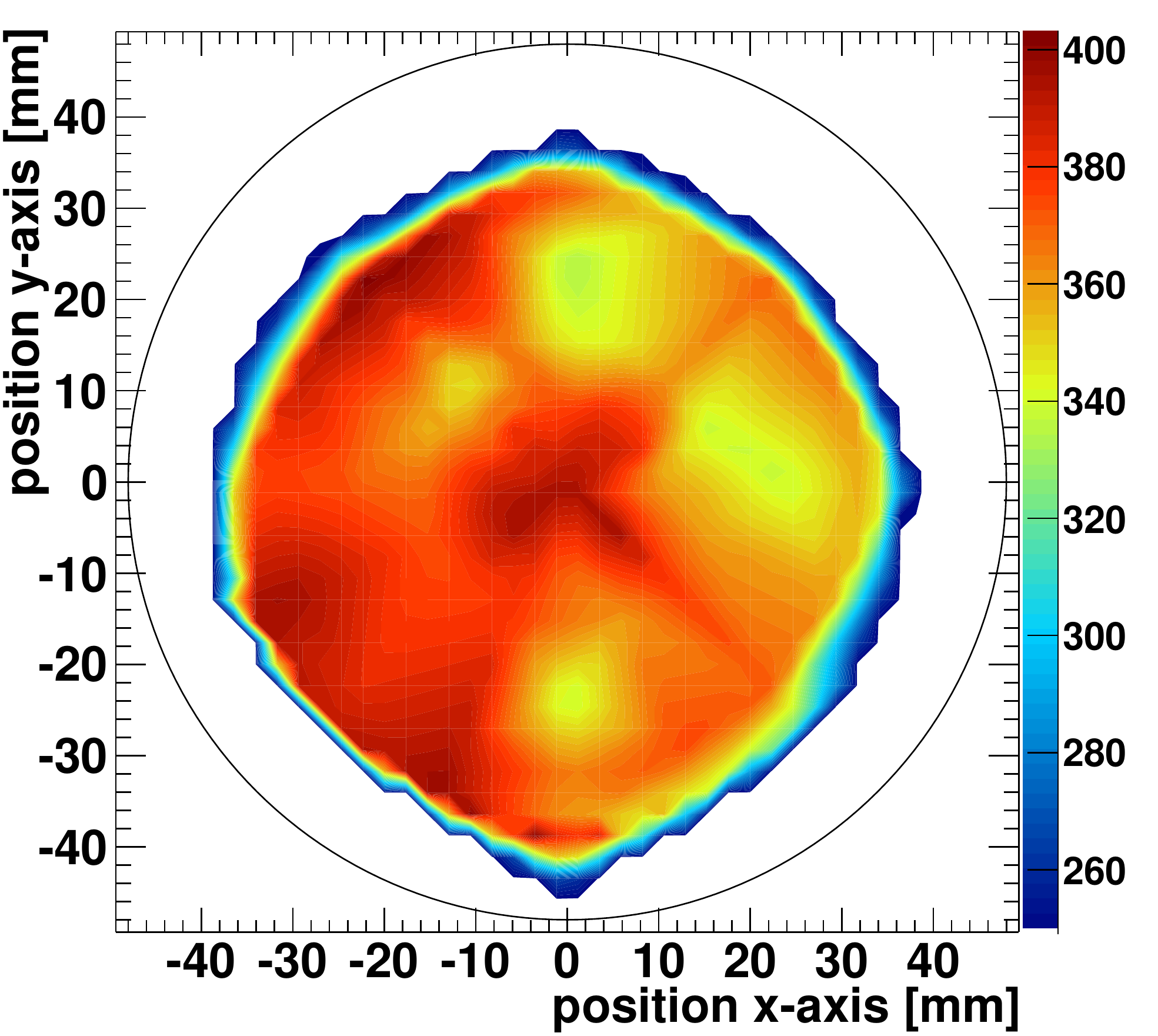}
\caption{Combined result of all surface scans on top of the diode. The color axis represents the measured count rates as a function of position. Also drawn is a circle which indicates the boundary of the Cu cryostat end cap~\cite{hakenmueller2012}.}
\label{fig:colorpic}
\end{center}
\end{minipage}
\end{figure*}

In a first step, fine-grained scans have been carried out with a collimated $400$\,kBq $^{241}$Am source to check the position of the diode inside the vacuum cryostat using a dedicated scanning table, originally designed for acceptance tests of the GERDA Phase II BEGe detectors~\cite{andreotti2013}. Horizontal scans along the top as well as vertical ones along the side of the diode housing are performed. As a result, it is found that the diode is slightly off-axis inside the cryostat as illustrated in Figure~\ref{fig:circle scans}. The figure displays the results of scans along circles of different radii on top of the diode inside the cryostat end cap. Concerning the scan along the maximum radius of $38$\,mm, that is still supposed to lie within the nominal diode boundary (green data points in Figure~\ref{fig:circle scans}), the count rate drops in a certain angular range between $0$ and $180^\circ$, indicating that the collimated $\gamma$ ray beam hits the diode only peripherally -- if at all -- at these positions. Figure~\ref{fig:colorpic} combines all available data to show a two dimensional representation of the measured surface variations on top of the diode. The circle line represents the cryostat boundary. The two dimensional map confirms the slightly asymmetric position of the diode. This has been considered in the implementation of the final detector geometry, as depicted in the right panel of Figure~\ref{fig:simulation of cryostat}.

In a second step, measurements to determine the dead-layer thickness and the active volume are carried out by using uncollimated $^{241}$Am, $^{133}$Ba, and $^{60}$Co sources. These sources were placed at different positions around the detector, and the number of counts for various $\gamma$~ray lines is measured. 

To evaluate the dead-layer thickness, the $^{241}$Am and $^{133}$Ba sources are employed because the $\gamma$ radiation of these sources does not penetrate the diode beyond a few mm. The intensity of the low-energetic lines is strongly affected by the dead-layer, while for high-energetic lines the influence is smaller. Thus the ratio of intensities between high-energetic and low-energetic lines serves as a measure for the dead-layer thickness. Using the ratio has the further advantage that uncertainties in the source activity, source distance and detector dead-time drop out.
All measurements are simulated with MaGe, assuming different dead-layer thicknesses. For each individual assumption the same ratio of $\gamma$ ray line intensities was calculated as done for the measurement. By comparing the ratios from MC and data, the dead-layer thickness is obtained as described in \cite{budjas2009}. According to this procedure the average dead-layer size amounts to ($1.5\pm0.1$)~mm. The obtained value is $\sim1$\,mm thicker than the nominal value. However, considering a storage time of the diode at room temperature for about $8$~years (see Section~\ref{sec:radiopurity}) the observed dead-layer growth is in agreement with $\sim$$0.1$\,mm/year observed also for other Ge detectors~\cite{huya2007,budjasheisel2009}. The dead-layer around the borehole cannot be determined in this way and is set to $0.3$\,$\mu \rm m$ in the simulations, according to information from the manufacturer.

Finally, the size of the active detection volume is determined by comparing the total measured count rate from $^{60}\rm Co$ exposure of the detector to results from the simulation. The $1173$\,keV and $1333$\,keV lines of the source are simulated for variable borehole dimensions keeping all other parameters fixed. The count rate from the simulation is plotted against the corresponding active volume and a linear fit is applied~\cite{budjas2009}. The comparison with measured data results in a best-fit active volume of ($339\pm23$)~$\rm cm^3$, which corresponds to an active mass of (1.81$\pm$0.12)\,kg.  
	
\subsection{Measurement and MC validation of the $\gamma$ ray detection efficiency}
\begin{figure}[tbh]
\includegraphics[width=1.0\columnwidth]{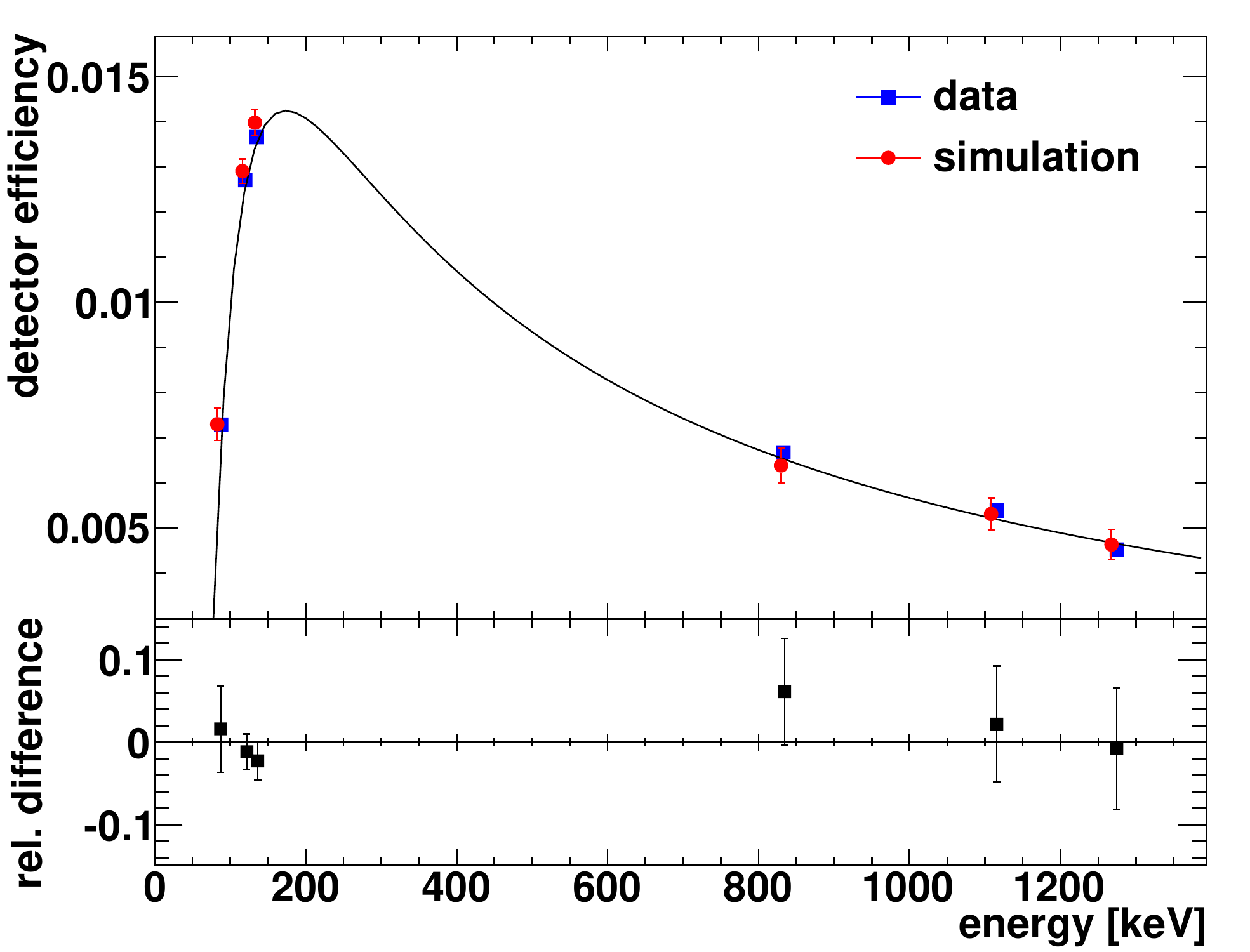}
\caption{Measurement and Geant4 simulation of the $\gamma$~ray detection efficiency as a function of energy in the GIOVE setup, determined by several monoenergetic sources at a distance of about $10$\,cm from the Ge detector end cap. Also shown is the relative deviation between simulation and data in the bottom panel.}
\label{fig:eff_curve}
\end{figure}

The detection efficiency, i.e. the probability to detect the full absorption energy of $\gamma$~rays emitted from an external source position, is determined by using several monoenergetic sources ($^{22}\rm Na$, $^{54}\rm Mn$, $^{57}\rm Co$, $^{65}\rm Zn$, $^{109}\rm Cd$) with activities in the range of about $10$\,kBq to $100$\,kBq. The measurements are all carried out at the same distance of about $10$\,cm to the cryostat end cap in a central position. The efficiency for each $\gamma$ energy line is computed as the ratio between the number of counts detected in the line and the number of $\gamma$ rays emitted at that energy. The latter is computed using the known calibration source activity and the corresponding decay branching ratio. The results for various absorption lines are plotted as blue data points in Figure~\ref{fig:eff_curve} and fitted with a polynomial logarithmic function divided by energy, which is commonly used to fit Ge spectrometer efficiency curves~\cite{debertin1984}.

To test the implementation of the Geant4 detector geometry (see Section~\ref{sec:mcmodel}) simulations with exactly the same configuration are carried out. The efficiencies predicted from the MC method are in excellent agreement with the calibration data -- as also shown in Figure~\ref{fig:eff_curve} (red points). The uncertainty of the simulated values contains systematic effects on the position and material of the source and on the size of the active Ge crystal volume. The latter uncertainty is caused by the fact that high-energetic $\gamma$ rays illuminate the whole active volume and are thus more influenced by variations in the bulk size than low energetic external $\gamma$ rays, which are detected predominantly in the superficial regions of the active volume. All effects combine to a total uncertainty estimate of about $3\%$. The mean absolute deviation between measurements and simulation of $(2 \pm 1)\%$ is similar to the agreement reached for other Ge detector setups at MPIK, such as Corrado~\cite{budjasheisel2009}. Altogether, these results indicate a successful implementation and optimization of the Ge detector geometry in the Geant4 MC code.

\section{Detector Performance and Screening Results}\label{sec:performance}
In this chapter we present results on the achieved low-background performance of the GIOVE spectrometer after finishing the construction of the detector and shield. It includes an analysis of the remaining background components, an evaluation of the achieved sensitivity, and a comparison to other instruments operated both at shallow depth and deep underground facilities. Preliminary results have been reported in~\cite{heusser2013}.     

\subsection{Spectral background analysis}
\label{subsection:giove-bckg-analysis}
\begin{figure}
\includegraphics[width=\columnwidth]{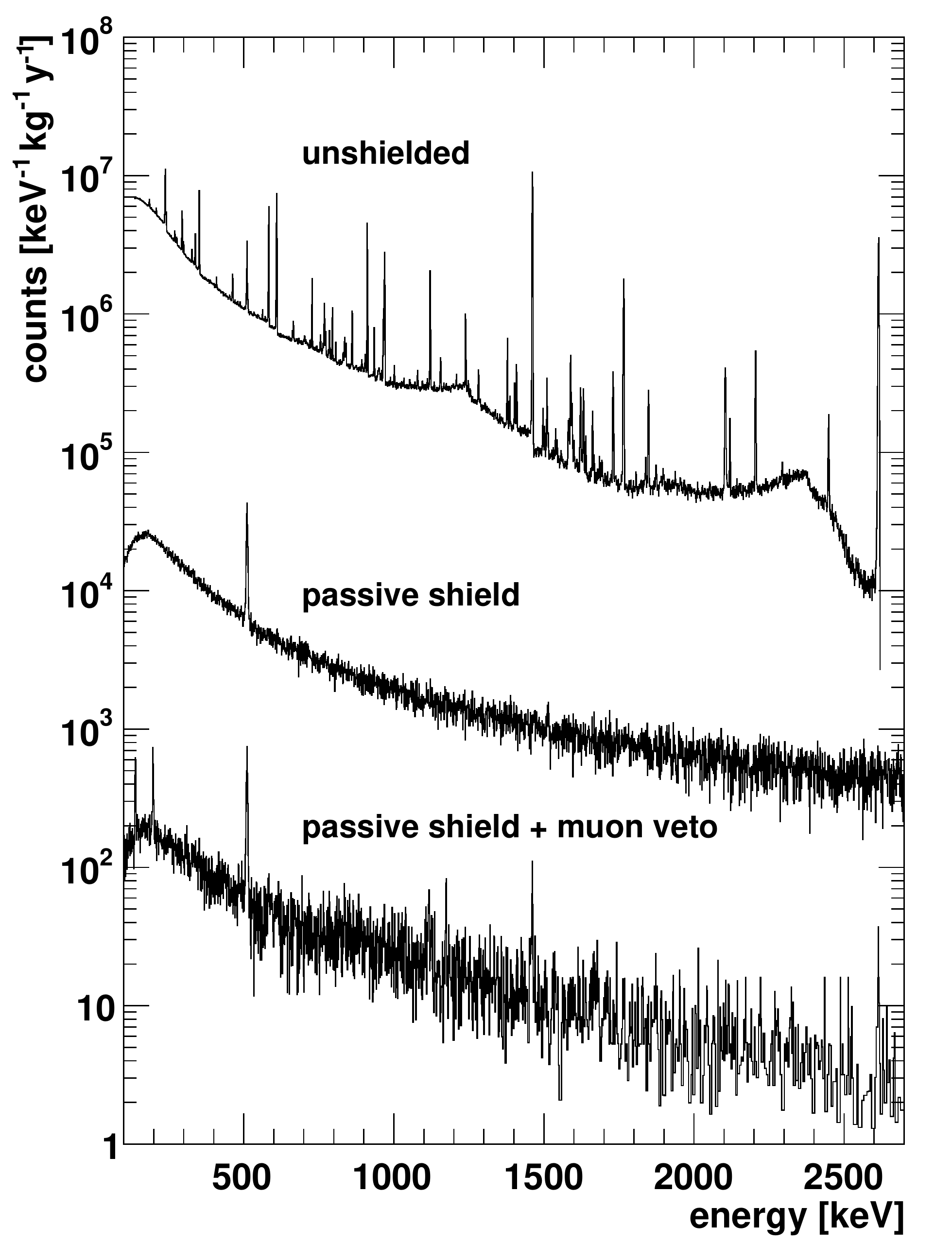}
\caption{Spectral background comparison of the GIOVE detector for increasing measures of passive and active shield.}
\label{fig:bgcomparisongiove}
\end{figure}

The overall reduction of the external radiation background measured by the GIOVE spectrometer is presented in Figure~\ref{fig:bgcomparisongiove} for different stages along the shield integration. The highest overall rate was recorded before the construction of the passive shield. The corresponding spectrum features a large variety of environmental $\gamma$~ray background lines, most prominently emerging from the decay of $^{232}$Th and $^{238}$U chain daughters, and $^{40}$K.
\begin{figure}[tbh]
\includegraphics[width=\columnwidth]{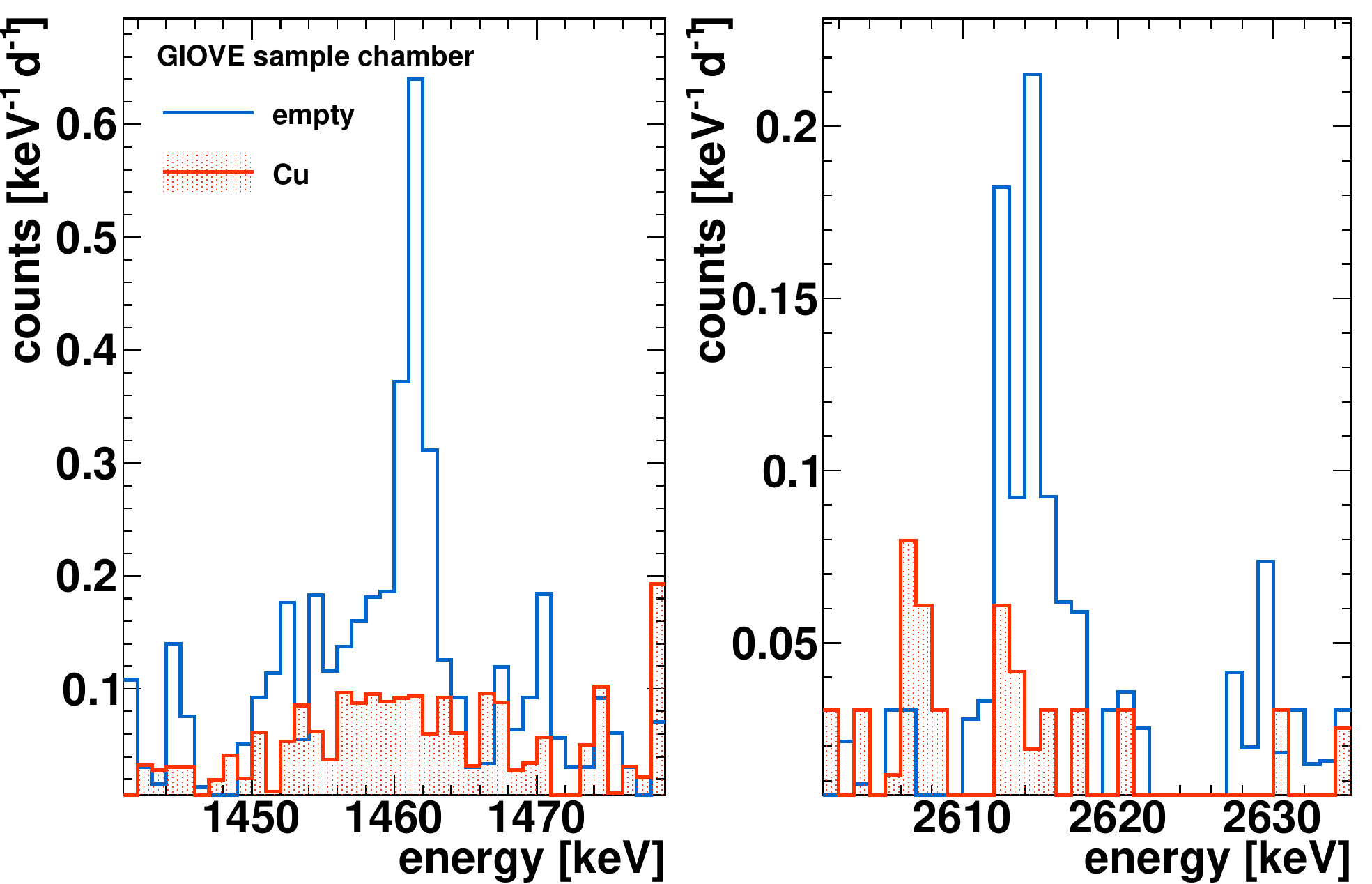}
\caption{Effect of additional Cu filling the GIOVE sample chamber (red histogram) on the two most prominent $\gamma$ ray line background regions at $^{40}$K (left) and $^{208}$Tl energies (right). Results indicate that the cause of remaining line backgrounds (blue histogram) is due to impurities outside of the cryostat end cap.}
\label{fig:linebgcomparisongiove}
\end{figure}
After completion of all passive shield layers the count rate is already suppressed by more than two orders of magnitude and most of the natural line contributions from external radioactivity have disappeared. The continuum and the strong presence of the $511$\,keV annihilation peak indicate that the spectrum is now dominated by muons and muon induced secondary particle showers ($\gamma$, e$^-$, e$^+$, neutrons).

The next step in total background reduction is \linebreak achieved by the application and optimization of the GIOVE active muon veto system. The spectrum comprises data collected in an uninterrupted long-term measurement with empty sample container over an effective live time of $33$ days, already corrected for its $\sim$$2\%$ dead-time loss. The integral count rate in the reference interval of ($40,2700$)~keV has diminished from about $39$\,min$^{-1}$ without veto to only ($0.44\pm0.01$)~min$^{-1}$, corresponding to a $89$ times reduction of the background when applying the anti-coincidence requirement with settings as described in Section~\ref{sec:muonveto}. The achieved suppression corresponds to a muon rejection efficiency at its designed value of $\sim$$99\%$.
At the same time as major parts of the prompt muon induced spectrum are successfully suppressed by the active veto, secondary but delayed components, such as neutron induced Ge isomer lines, emerge from the remaining background continuum.

In order to understand the origin and composition of the remaining background counts, the empty sample chamber measurement is compared to two further measurements, in which the sample chamber was either filled with $65.2$\,kg of high purity Cu ($32.8$\,d live time) or B loaded paraffin wax ($13.3$\,d). In all three cases the $\gamma$ ray line intensities of prominent radioisotopes/isomers in the recorded spectra are studied. For the definition of regions of interest (ROI) around the peaks and for the subtraction of background the DIN standard 25482-5 \cite{din-norm} is applied. Besides the statistical uncertainty, systematic contributions arising from the calibration of peak positions and full-widths at half-maximum are considered in the total uncertainty budget. For finite results a coverage factor k=1 (68\% confidence level C.L.) is quoted, otherwise a 90\% (k=1.645) C.L. is chosen to define upper limits according to \cite{heisel2009}. The results on the specific count rates, normalized in terms of counts per day, are provided in Table~\ref{tab:lineresults}.

In the case of the empty sample chamber (third column) we observe small $\gamma$ line contributions from the radioactive isotopes $^{40}$K, $^{60}$Co and from high energetic 2.6\,MeV $\gamma$ rays emitted in decays of the $^{228}$Th daughter nuclide $^{208}$Tl. The rates are in the range of $1.8$, $1.0$ and $0.5$\,d$^{-1}$, respectively. The low count rates in the $^{226}$Ra ($^{222}$Rn) progeny lines demonstrate the effectiveness of the Rn suppression system. If the empty sample chamber is no longer flushed with N$_2$ and ambient air with a typical $^{222}$Rn concentration of about 60\,Bq\,m$^{-3}$ is allowed to enter through the opened glovebox, the intensities of the corresponding $\gamma$ ray lines rise by more than two orders of magnitude. On the contrary, if all relevant provisions described in Section~\ref{sec:designprinciples} are taken, the total count rate is immediately at the value quoted in Table~\ref{tab:lineresults} after the sample insertion. The Ge isomer lines at 53, 139 and 198\,keV as well as the muonic annihilation peak at 511\,keV have intensities of around $7$, $9$, $9$ and $19$\,d$^{-1}$.  

The fourth column shows the evaluation of the measurement in which the Cu block was placed inside the sample chamber surrounding the cryostat end cap. Besides a notable reduction of the integral background rate, the $^{40}$K and $^{208}$Tl lines are strongly suppressed -- as depicted in Figure~\ref{fig:linebgcomparisongiove} -- and only upper limits can be found. The outcome clearly indicates that the signals seen in the empty chamber measurement are likely caused by remaining contaminations of outer shield components or from external environmental radiation. Consequently, they do not originate from the cryostat end cap or from the Ge detector themselves. In the case of $^{60}$Co, the rate drop from $1.0$ to $0.5$\,d$^{-1}$  indicates that this contaminant is likely to be located to some extent outside the cryostat end cap. We also learn that line intensities coming from the de-excitation of the neutron induced isomeric Ge states are hardly altered by the presence of Cu, because its effect on the neutron attenuation length is comparatively small.

The fifth column summarizes the count rates obtained after deployment of a B doped paraffin wax block inside the sample chamber. While some of the neutron induced metastable Ge isotopes reach a suppression factor of $\sim$2, the overall contribution from natural radioactivity increases by up to a factor of $\sim$3. As found in subsequent measurements, other hydrogen-rich neutron absorbers behave similarly, preventing their application in close vicinity of the cryostat end cap as potential shield materials. In this way, the measurement reported here demonstrates one of the remaining challenges in the construction of an optimal passive shield for a Ge detector operated at shallow depth.
\begin{table*}[t] \small
\caption[]{Count rates of $\gamma$~ray lines from natural and anthropogenic radioactivity as well as induced by muons and muonic by-products. The rates obtained in different sample chamber (SC) configurations are reported. Mean $\gamma$ ray line energies are listed according to~\cite{recommendeddata}.}
\label{tab:lineresults}
		\centering
		\begin{tabular*}{1.0\textwidth}[tbh]{@{\extracolsep{\fill}}llccc}
\hline 									
isotope/isomer&energy [keV]&	SC: empty [d$^{-1}$] 		&SC: 65.2\,kg Cu [d$^{-1}$]		&SC: B doped paraffin [d$^{-1}$]\\
\hline																										
$^{226}$Ra	&	295.22	&	$<$1.3		&	$<$1.0		&	$<$2.0		\\
			&	351.93	&	1.2$\pm$0.5	&	1.3$\pm$0.4	&	2.1$\pm$0.6	\\
			&	609.32	&	$<$0.6		&	$<$1.1		&	3.1$\pm$0.5	\\
$^{238}$U	&	1001.44	&	$<$0.5		&	$<$0.3		&	$<$0.4		\\
$^{228}$Th	&	238.63	&	$<$1.4		&	$<$0.9		&	2.4$\pm$0.8	\\
			&	583.19	&	$<$0.5		&	$<$0.9		&	2.0$\pm$0.5	\\
			&	2614.51	&	0.5$\pm$0.2	&	$<$0.3		&	1.7$\pm$0.3	\\
$^{228}$Ra	&	911.21	&	$<$0.5		&	0.8$\pm$0.2	&	1.5$\pm$0.4	\\
			&	964.77+968.97	&	$<$0.6		&	$<$0.6		&	2.0$\pm$0.4	\\
$^{40}$K		&	1460.82	&	1.8$\pm$0.3	&	$<$0.4		&	3.9$\pm$0.5	\\
$^{137}$Cs	&	661.66	&	$<$0.6		&	$<$0.2		&	0.8$\pm$0.4	\\
$^{60}$Co	&	1172.24	&	1.0$\pm$0.3	&	0.6$\pm$0.2	&	0.4$\pm$0.3	\\
	        &	1332.51	&	1.1$\pm$0.3	&	0.5$\pm$0.2	&	0.4$\pm$0.2   \\
$^{73m}$Ge	&	53.5		&	6.6$\pm$1.1	&	6.4$\pm$1.0	&	5.3$\pm$1.1	\\
$^{75m}$Ge	&	139.5	&	8.8$\pm$0.8	&	10.0$\pm$0.7	&	5.5$\pm$1.0	\\
$^{71m}$Ge	&	198.3	&	9.3$\pm$0.8	&	9.6$\pm$0.7	&	5.5$\pm$0.9	\\
e$^+$e$^-$ annih.& 511.00 &	19.3$\pm$1.1	&	17.1$\pm$0.8	&	15.3$\pm$0.9	\\
\hline			
		\end{tabular*}
\end{table*}

\subsection{Achieved sensitivity and first $\gamma$ ray screening measurements}
\label{subsection:giove-sensitivity}

\begin{table*}[tb] \small
  \caption[]{Achievable sensitivities with the GIOVE detector for 30 days counting time. For two classes of samples the threshold activity for various radioactive isotopes is calculated. For class 1 samples the achievable sensitivity can vary between the two extreme cases given in the Table due to the self shielding effect. See text for further explanations.}
\label{tab:decision_thresholds}
		\centering
		\begin{tabular*}{1.0\textwidth}[tbh]{@{\extracolsep{\fill}}cccc}
\hline 	
& \multicolumn{2}{c}{Class 1: Small low-mass samples} & Class 2: Large high-density samples \\
& \multicolumn{2}{c}{represented by getter stripe}  &  represented by copper block \\ \hline
Sample chamber configuration & SC: empty & SC: 65.2\,kg Cu &  SC: 65.2\,kg Cu\\ \hline
Isotope  & Threshold      &  Threshold      & Massic threshold \\
         & activity [mBq] &  activity [mBq] & activity  [mBq\,kg$^{-1}$]	\\
\hline												
$^{238}$U    & 13     &  11    &  4     \\
$^{226}$Ra   & 0.20   &  0.19  &  0.07  \\
$^{228}$Ra   & 0.4    &  0.4   &  0.11  \\
$^{228}$Th   & 0.4    &  0.29  &  0.05  \\
$^{40}$K     & 1.8    &  0.9   &  0.24  \\
$^{60}$Co    & 0.15   &  0.12  &  0.026 \\
$^{137}$Cs   & 0.12   &  0.09  &  0.04  \\
\hline	
\end{tabular*}
\end{table*}

In this section we determine first the benchmark sensitivity that can be obtained with the GIOVE detector in measuring smallest activity concentrations of primordial, anthropogenic and cosmogenic decay isotopes in materials. Then we report on examples of materials that have been screened with GIOVE.\\

For the sensitivity determination we evaluate some of the background measurements presented in the previous section in a different way. This time we are not interested in the deconvolution of the different background components for a given $\gamma$ ray line. Instead, we want to estimate the benchmark sensitivity by considering all recorded events in that region of interest (ROI) as the expectation value for the background contribution in a subsequent sample measurement. From this expectation value we calculate the decision threshold (DT) for the corresponding $\gamma$ ray line according to \cite{din-norm}. It is clear that the achievable sensitivity depends on the counting time. In order to compare measurements of different duration we normalize all results to a typical standard counting time of 30 days.

In order to convert the numerical value of DT (in counts) into a threshold activity  (in mBq) we need the branching ratio and the total counting efficiency of the $\gamma$ ray line under investigation. The latter one depends strongly on the properties of the sample, thus a generic number cannot be given. For the benchmark sensitivity estimation we consider two reference samples: A 43\,mg getter stripe represents the class of small low-mass samples which can be placed close to the HPGe crystal. The second class encompasses large volume and high-density samples, which fill almost the entire sample chamber. They are represented by the already mentioned 65.2\,kg sample of electrolytic copper which was used for the background study in section \ref{subsection:giove-bckg-analysis}. For both classes the counting efficiencies are simulated (Section~\ref{sec:mcmodel}) and the threshold activities (in mBq) are calculated for each ROI. Whenever several $\gamma$ ray lines belong to the same sub-chain in radioactive equilibrium they are combined using the weighted average to obtain the combined threshold activity as the best estimate for the overall sensitivity.  Results for the second class of samples are normalized to the mass of the copper sample.

As seen in the previous section, part of the residual background radiation is located outside of the sample chamber.  Depending on the size, composition and density of the samples this fraction of the background radiation may be suppressed by the self shielding effect. For the second class of samples complete self shielding is always provided in very good approximation. For the first class of samples this is not the case, so we quote the sensitivity without self-shielding (empty sample chamber) and with strong self-shielding (65.2 kg of copper) as the two extreme cases. The results are given in Table \ref{tab:decision_thresholds}.

The absolute threshold activity for class 1 samples lies slightly above 100 $\mu$Bq for $^{60}$Co and $^{137}$Cs and in the range of few hundred $\mu$Bq for $^{226}$Ra, $^{228}$Ra and $^{228}$Th.  It can be seen that an additional copper shield in the sample chamber improves the numbers only a bit, indicating that the chosen passive shield is well dimensioned and has a high radiopurity. The only exception is $^{40}$K for which the threshold activity is about a factor of two lower if copper is present ($1.8$\,mBq versus $0.9$\,mBq). For class 2 samples the sensitivity for $^{226}$Ra, $^{228}$Ra and $^{228}$Th is in the range of 50 to 100 $\mu$Bq\,kg$^{-1}$ and around 30 $\mu$Bq\,kg$^{-1}$ for $^{60}$Co and $^{137}$Cs.  It is expected that the quoted threshold activity for $^{60}$Co is a conservative number as copper contains traces of $^{60}$Co due to cosmogenic activation. For $^{40}$K the threshold activity is 240 $\mu$Bq\,kg$^{-1}$. To summarize we can state that the GIOVE design specifications have been fully accomplished and in particular a sensitivity of $\leq$ 100 $\mu$Bq\,kg$^{-1}$ for $^{226}$Ra and $^{228}$Th has been reached.\\

From the material samples that have been screened with GIOVE so far we will here present the results of three representative examples: The getter stripe and the copper block, which have already been mentioned above, and a 36.2 kg stainless steel sample. All results are summarized in Table \ref{tab:three_results}.

The already introduced 43\,mg getter stripe is an active zirconium alloy deposited on a metallic substrate. The getter is used as a component inside a \textsc{Hamamatsu} PMT of type R11410-21 which will be employed in the upcoming XENON1T dark matter experiment. For screening it is placed directly on the cryostat end cap and surrounded with the previously discussed Cu block as an additional shield. The screening measurement lasted 19.3\,d. For most radionuclides only upper limits can be found, but a notable contamination of ($2.6\pm0.9$)\,mBq for $^{40}$K is detectable.

For the evaluation of the activity of the copper sample itself we use the same spectrum as for the calculation of the threshold activity, but now we process it in the standard way by estimating the continuous background in each ROI from the neighbouring channels. Any observable excess would then be induced by radioactivity in the copper. Due to the efficient self-shielding effect no line background is subtracted, but we note that some contamination may still sit in the cryostat and cannot be (self-)shielded.

\begin{table}[tb] \small
\begin{minipage}[t]{1.0\columnwidth}
\caption{Results of three representative samples screened with GIOVE.}
\label{tab:three_results}
\begin{tabular*}{1.0\columnwidth}[tbh]{@{\extracolsep{\fill}}cccc}
\hline 	
Sample & Getter stripe & Cu block & Stainless steel\\
Mass & 0.043 g & 65.2 kg & 36.2 kg \\ \hline
Isotope & Activity & Massic activity & Massic activity \\
  & [mBq] &  [mBq\,kg$^{-1}$] &  [mBq\,kg$^{-1}$] \\
\hline												
$^{238}$U	& $<$13.6       	& $<$2.5         	&$<$15.8     \\
$^{226}$Ra	& $<$0.32       	& 0.13$\pm$0.04  	&$<$0.43     \\
$^{228}$Ra	& $<$0.53       	& 0.14$\pm$0.06  	&$<$0.53     \\
$^{228}$Th	& $<$0.45       	& $<$0.06         	&$<$0.16     \\
$^{40}$K		& 2.6$\pm$0.9  	& $<$0.26         	&1.9$\pm$0.5 \\
$^{60}$Co	& $<$0.20       	& 0.04$\pm$0.01  	&7.9$\pm$0.4 \\
$^{137}$Cs	& $<$0.12       	& $<$0.02         	&$<$0.24     \\
\hline
\end{tabular*}
\end{minipage}
\end{table}

Finally, we re-measured 36.2\,kg of a stainless steel sheet that was used for the construction of the cryostat of the GERDA experiment. The same material has already been screened with Dario, one of the three first generation spectrometers operated at the MPIK \cite{aberle2010}. It is denoted as sample D6 in Ref.~\cite{maneschg2008}, in which the old results are published.  The new GIOVE measurement significantly improves these results in a comparable screening time illustrating its superior performance. In particular the limits for the $^{228}$Ra and $^{228}$Th concentrations of $<$1.4 and $<$0.8\,mBq\,kg$^{-1}$ are lowered to $<$0.53 and $<$0.16 mBq\,kg$^{-1}$, respectively.

\subsection{Benchmark comparison}
\begin{table*}[tb] \small
\caption[]{Integral count rates of low-background Ge spectrometers located in different underground sites. The values are normalized by the active mass (m$_{act}$) of the detectors. Among them, only GIOVE and Corrado are equipped with active muon veto systems. The sample chambers (SC) were empty during the measurement.}
		\label{tab:bkgr-benchmark}
		\centering
		\begin{tabular*}{1.0\textwidth}[tbh]{@{\extracolsep{\fill}}llllll}
\hline 									
detector	&	m$_{act}$ 	&	SC$_{vol}$	&	location, depth &	$\mu$ flux reduction	&	count rate in (40,2700)\,keV 	\\
		&	[kg]		&	[liter]	&	[m w.e.]			&(comp. sea level)	&	[d$^{-1}$kg$^{-1}$]		\\
\hline											
\vspace{0.2cm}
GIOVE	&	1.81		&	12.4		&	MPIK, 15 		&	2$-$3			&	348$\pm$3	\\
Corrado	&	0.94		&	11		&	MPIK, 15 		&	2$-$3			&	3661$\pm$11	\\
Ge-3		&	1.24		&	0.4		&	HADES, 500		&	5$\times$10$^3$	&	394$\pm$2	\\
GeMPI	&	2.06		&	15		&	LNGS, 3800		&	10$^6$			&	66$\pm$1	\\
\hline		
		\end{tabular*}
\end{table*}

In the last section we compare the overall background levels achieved with the GIOVE setup and other existing spectrometers, located at various underground sites and using different shield designs.

\begin{figure*}[tb]
\center
\includegraphics[width=1.0\textwidth]{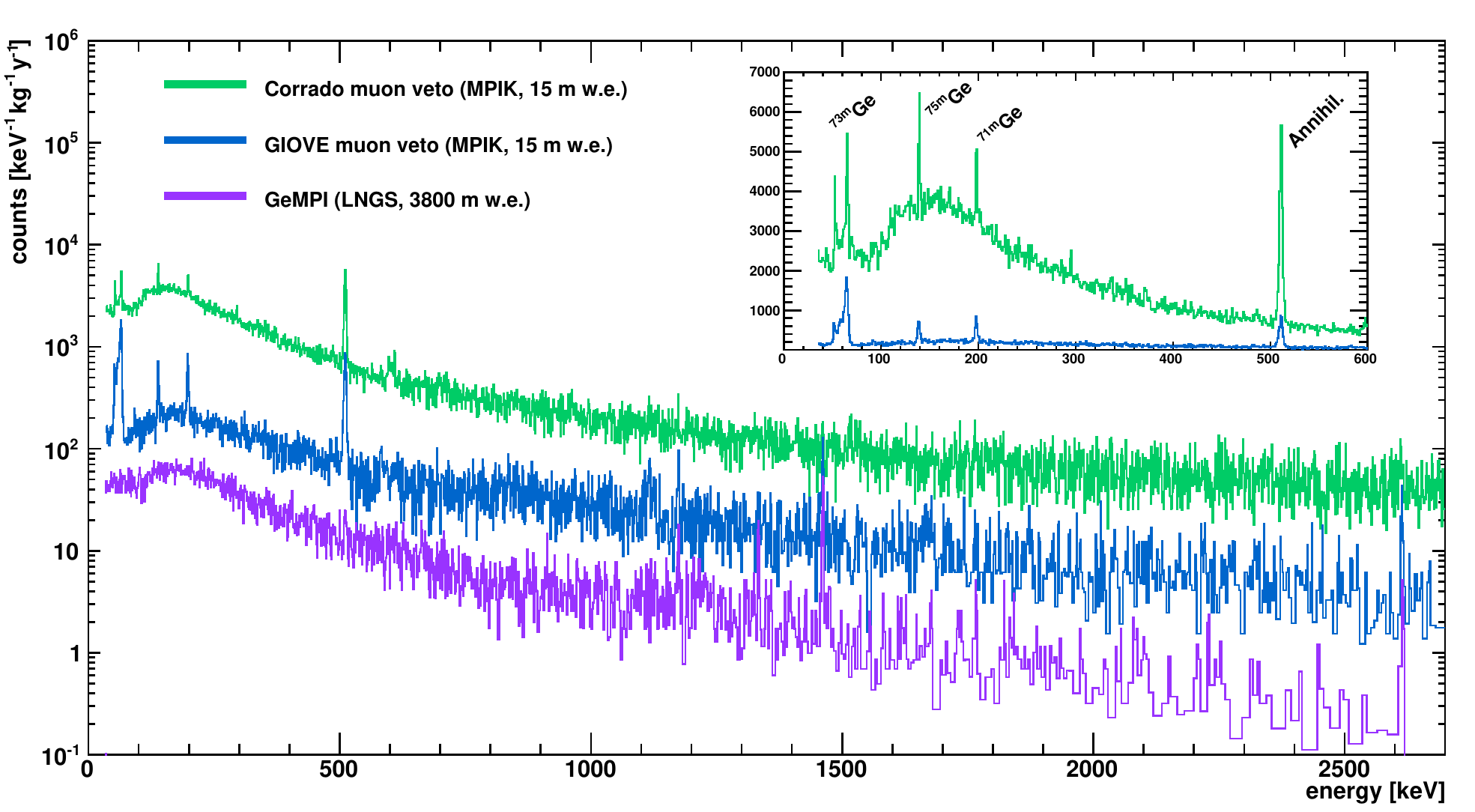}
\caption{Comparison of long-term $\gamma$ ray background measurements between spectrometers employing various shield concepts at different laboratory depths. Energy spectra in the most relevant range of $\gamma$ ray screening for natural radioactivity within ($40,2700$)\,keV are depicted for the MPIK spectrometers GIOVE and Corrado, and the passive shield GeMPI device at the LNGS deep underground facility.}
\label{fig:bgcomparisonglobal}
\end{figure*}

One of the initial objectives of the GIOVE design was to achieve a significant reduction in the muon induced and radio-impurity related background rate compared to the previous generation of Ge spectrometers at the MPIK underground laboratory, such as Corrado~\cite{budjas2008}. The latter has a comparable Pb and Cu shield thickness but features only an outer shell active muon veto system, which consists of several proportional gas ionization wire chambers (coverage from all sides except from below) \cite{maneschg2007}. Moreover, there are no neutron absorptive layers implemented.

A comparison of two long time background measurements with both detectors, each applying its own muon anti-coincidence system, is included in Figure~\ref{fig:bgcomparisonglobal}. All rate spectra in this plot refer to empty sample chamber measurements and are given in units of keV$^{-1}$\,kg$^{-1}$\,y$^{-1}$, i.e. normalized to the respective active Ge crystal mass. The improvement of the GIOVE background rate observed in the low energy region (linear scale plot embedded in Figure~\ref{fig:bgcomparisonglobal}) is coherent up to highest energies relevant for natural $\gamma$ ray screening as shown in the full scale plot of Figure~\ref{fig:bgcomparisonglobal}. As quantified in Table~\ref{tab:bkgr-benchmark}, the integral background rate in the (40,2700)\,keV window is reduced by more than a factor of ten to 348\,d$^{-1}$kg$^{-1}$, mainly by increasing the tagging efficiency of muons and thus improving the veto performance for prompt and delayed muon induced signals. Thus, the aimed muonic background suppression has been successfully achieved. In addition, the absolute rates of the neutron activated lines are measured to be $\sim70\%$ lower in the GIOVE setup than in Corrado. The observed suppression factor is therefore in good agreement with the expected drop in signal intensity derived from the initial neutron attenuation experiment presented in Section~\ref{sec:neutronattenuation}.

In order to evaluate an equivalent (``virtual'') depth effected by the GIOVE muon veto system, we compare the shield performance with low-background HPGe detectors operated without active muon vetos at moderately deep underground sites. Ge-3, for instance, is a 1.2\,kg semi-coaxial HPGe detector located at the\linebreak HADES underground laboratory~\cite{hult2000}. The overburden of 500~m~w.e. reduces the muon flux by a factor of $\sim$5000 compared to sea level \cite{andreotti2011}. At this depth and beyond, contributions from natural radioactivity in shield materials or ambient air start dominating over those from muons and muon induced by-products (see Figure 1 in \cite{laubenstein2004}). The tight passive shield used for Ge-3 consists of 14\,cm of Pb, low in $^{210}$Pb, and 14\,cm of electrolytic Cu. The obtained integral count rate for this detector configuration is 1.1 times higher than the one obtained by GIOVE. An investigation of the low energy part of the spectrum reveals -- as expected -- that longer delayed Ge isomer de-excitations observed at shallow depths (GIOVE, Corrado) are not present in Ge-3~\cite{hult2014}. Most of the other $\sim$10 low-background HPGe detectors which are operated in HADES and embedded in different Pb-Cu shield configurations have integral count rates above $\sim$400\,d$^{-1}$kg$^{-1}$ and only in few cases around $\sim$300\,d$^{-1}$kg$^{-1}$ \cite{hult2013}. The best detection limits for U and Th reached with Ge detectors in HADES are in the range of $100$\,$\mu$Bq\,kg$^{-1}$, very similar to GIOVE.

Finally we compare GIOVE with the GeMPI detectors at the LNGS underground facility~\cite{heusser2006,neder2000}. At their location, underneath 3800\,m~w.e. of rock overburden, the muon flux is suppressed by six orders of magnitude. Similar to the HADES spectrometers, these low-background HPGe detectors do not need to be equipped with an active muon veto. Their background rate level is mostly limited by the finite radio-purity of the Ge crystal and surrounding shield materials. A 102\,d lasting GeMPI measurement with empty sample chamber led to an integral count rate of 66~d$^{-1}$kg$^{-1}$, which is $\sim$5 times below the one obtained with GIOVE.

\section{Summary and Outlook}\label{sec:summary}
GIOVE, a new Ge spectrometer equipped with a passive shield consisting of carefully selected radio-pure materials and a newly developed active muon veto system, has been successfully built at our shallow depth underground laboratory (15\,m~w.e.).

In a series of background studies we demonstrate that our original design specifications for the new setup are met. We obtain a rejection efficiency of prompt signals caused by muons and muon induced by-products of $\sim$$99$\% with an acceptable dead-time fraction of $<2$\%. Further, the integration of neutron-absorbing layers allows us to reduce neutron induced signals by $\sim$70\%. In this way, we are able to reach integral count rates that are typically obtained at moderately deep underground sites of several 100\,m w.e. At such depths, the muon flux is already strongly suppressed and natural radioactivity from the environment and construction materials start to compete with the residual muon induced signals.

In terms of sensitivity for detecting primordial U and Th traces in $\gamma$ ray screening measurements we reach a level transcending $\sim$$100$\,$\mu$Bq\,kg$^{-1}$, which closes the sensitivity gap between currently existing low background Ge spectrometers operated at the same depth ($1$\,mBq\,kg$^{-1}$) and in deep underground locations \linebreak ($10$\,$\mu$Bq\,kg$^{-1}$). Since this sensitivity is high enough for screening a large fraction of materials for recent or future physics experiments, the installation of new GIOVE-like spectrometers at shallow depth would take off load from sparsely available screening facilities in deep underground locations, reducing long waits and transportation costs and giving the advantage of easy and more flexible access.\\
Besides advancements in simplifying material screening, the background levels achieved with the GIOVE setup are attractive for physics research programs which need low-background conditions but have to be carried out at shallow depths, e.g. beneath nuclear reactor cores or accelerators close to the surface.

In order to further optimize the novel shield design, we plan to perform MC simulations of GIOVE to reproduce the experimental results and understand in detail the creation, propagation and suppression of radiation background generated in a heavy element shield. Especially an even more effective moderation and absorption of neutrons will be of interest in order to reduce delayed neutron induced signals from de-excitations of the Ge isomers $^{71m}$Ge, $^{73m}$Ge and $^{75m}$Ge, but also from activated long-lived isotopes such as $^{68}$Ge and $^{60}$Co. The already established and validated code can be used to model optimal shield layer combinations as a function of depth and of available radio-purity of shield components to predict the background suppression efficiency of future instruments both at shallow and deep underground sites. 

\paragraph{Acknowledgements}
\footnotesize{We are deeply grateful to the technical engineering staff at the MPIK, in particular to Ralf Lackner and Michael Reissfelder, for their permanent support in the construction of GIOVE. We would like to thank Bettina M\"ork for the preparation of technical drawings and help in the mechanical design. We also appreciate the help from the electronics workshop at MPIK for the manufacture of a custom designed muon trigger logic unit. Stefan Sch\"onert is thanked for the coordination of the crystal fabrication within a record short surface exposure of one week as part of the GERDA experiment. Furthermore, we are very thankful for the many fruitful discussions with Alan W. P. Poon and his support with MC simulations to optimize the neutron flux suppression in the early construction phase of GIOVE. We also thank Florian Kaether for his valuable support in the development of the formalism for sensitivity calculation. We are indebted to Mikael Hult for hosting the GIOVE crystal at HADES and organizing many transfers from there to Canberra, Olen.}

\bibliographystyle{unsrt}

\end{document}